\documentclass[aoas]{imsart}
\RequirePackage{amsthm,amsmath,amsfonts,amssymb}
\RequirePackage[authoryear]{natbib}

\startlocaldefs
\usepackage{cancel}
\usepackage{graphics}
\usepackage{color}
\usepackage{xcolor}
\usepackage{soul}
\usepackage{mathrsfs}
\usepackage{multirow}
\usepackage{hhline}
\usepackage{subcaption}
\usepackage{graphicx}
\usepackage[colorlinks,bookmarksopen,bookmarksnumbered,citecolor=blue,urlcolor=blue,linkcolor=blue]{hyperref}
\usepackage[capitalise,nameinlink]{cleveref}
\usepackage{lipsum}
\usepackage{float}
\usepackage{url}
\usepackage[ruled]{algorithm2e}
\usepackage{tikz}
\usepackage{booktabs}
\usepackage{longtable}
\usepackage{array}
\usepackage{xcolor}
\usepackage{colortbl}
\usepackage{hyperref}

\usepackage{xr}
\makeatletter
\newcommand*{\addFileDependency}[1]{
\typeout{(#1)}
\@addtofilelist{#1}
\IfFileExists{#1}{}{\typeout{No file #1.}}
}\makeatother

\newcommand{\mb}[1]{\mathbf{#1}}

\mathchardef\mhyphen="2D

\newcommand{\bbT}{\mathbb{T}}
\newcommand{\bbM}{\mathbb{M}}
\newcommand{\bbI}{\mathbb{I}}

\def\trans{^\mathsf{T}}

\newcommand{\sB}{\mathcal{B}}

\newcommand{\sN}{\mathcal{N}}

\newcommand{\sM}{\mathcal{M}}
\newcommand{\sD}{\mathcal{D}}

\newcommand{\vb}{\mathbf{b}}

\newcommand{\ve}{\mathbf{e}}
\newcommand{\vf}{\mathbf{f}}
\newcommand{\vg}{\mathbf{g}}
\newcommand{\vh}{\mathbf{h}}

\newcommand{\vr}{\mathbf{r}}
\newcommand{\vs}{\mathbf{s}}

\newcommand{\vx}{\mathbf{x}}

\newcommand{\vz}{\mathbf{z}}

\newcommand{\vpsi}{\boldsymbol{\psi}}
\newcommand{\vphi}{\boldsymbol{\phi}}

\newcommand{\vmu}{\boldsymbol{\mu}}

\newcommand{\mA}{\mathbf{A}}
\newcommand{\mB}{\mathbf{B}}

\newcommand{\mF}{\mathbf{F}}

\newcommand{\mI}{\mathbf{I}}

\newcommand{\mL}{\mathbf{L}}

\newcommand{\mS}{\mathbf{S}}
\newcommand{\mT}{\mathbf{T}}

\newcommand{\mV}{\mathbf{V}}

\newcommand{\mOmega}{\boldsymbol{\Omega}}

\newcommand{\mSigma}{\boldsymbol{\Sigma}}

\newcommand{\mPsi}{\boldsymbol{\Psi}}

\newcommand{\R}{\mathbb{R}}
\newcommand{\E}{\mathbb{E}}

\crefname{condition}{Condition}{Condtions}

\endlocaldefs

\begin{document}

\begin{frontmatter}
\title{How do the professional players select their shot locations? An analysis of Field Goal Attempts via Bayesian Additive Regression Trees}
\runtitle{Field Goal Attempts via Bayesian Additive Regression Trees}

\begin{aug}
\author[A]{\fnms{Jiahao}~\snm{Cao} \ead[label=e1]{jiahao.cao@uth.tmc.edu}\orcid{0009-0008-3850-178X}},
\author[B]{\fnms{Hou-Cheng}~\snm{Yang}\ead[label=e2]{houchengyang4343@gmail.com}\orcid{0000-0000-0000-0000}}
\and
\author[A]{\fnms{Guanyu}~\snm{Hu} \ead[label=e3]{guanyu.hu@uth.tmc.edu}\orcid{0000-0000-0000-0000}}

\address[A]{Center for Spatial Temporal Modeling for Applications in Population Sciences, Department of Biostatistics and Data Science, The University of Texas Health Science Center at Houston, Texas, USA\printead[presep={ ,\ }]{e1,e3}}

\address[B]{Department of Statistics, Florida State University, Florida, USA\printead[presep={ ,\ }]{e2}}
\end{aug}

\begin{abstract}
Basketball analytics has significantly advanced our understanding of the game, with shot selection emerging as a critical factor in both individual and team performance. With the advent of player tracking technologies, a wealth of granular data on shot attempts has become available, enabling a deeper analysis of shooting behavior. However, modeling shot selection presents unique challenges due to the spatial and contextual complexities influencing shooting decisions. This paper introduces a novel approach to the analysis of basketball shot data, focusing on the spatial distribution of shot attempts, also known as intensity surfaces. We model these intensity surfaces using a Functional Bayesian Additive Regression Trees (FBART) framework, which allows for flexible, nonparametric regression, and uncertainty quantification while addressing the nonlinearity and nonstationarity inherent in shot selection patterns to provide a more accurate representation of the factors driving player performance; we further propose the Adaptive Functional Bayesian Additive Regression Trees (AFBART) model, which builds on FBART by introducing adaptive basis functions for improved computational efficiency and model fit. 
AFBART is particularly well suited for the analysis of two-dimensional shot intensity surfaces and provides a robust tool for uncovering latent patterns in shooting behavior. Through simulation studies and real-world applications to NBA player data, we demonstrate the effectiveness of the model in quantifying shooting tendencies, improving performance predictions, and informing strategic decisions for coaches, players, and team managers. This work represents a significant step forward in the statistical modeling of basketball shot selection and its applications in optimizing game strategies.
\end{abstract}


\end{frontmatter}

\section{Introduction}
Basketball analytics has transformed our understanding of the game, uncovering insights that go well beyond conventional metrics like points scored and rebounds collected. Among the many elements of the game, shot selection stands out as a pivotal factor influencing both individual performance and team success. Knowing where and how players choose to shoot is invaluable for optimizing strategies, refining player development, and gaining a competitive edge. Yet, analyzing shot selection is a complex task due to the interplay of spatial and contextual factors that shape shooting decisions.

The advent of advanced player tracking technologies \citep{albert2017handbook} has revolutionized data collection in professional basketball, providing granular details on every shot attempt. These data capture not only the precise location of each shot but also contextual information, such as defensive pressure, player movement, and game dynamics. Such rich datasets have fueled the development of sophisticated statistical models aimed at deciphering the intricate patterns of shot selection \citep{cervone2014pointwise, franks2015characterizing, cervone2016multiresolution, wu2018modeling, hu2021cjs, fernandez2018wide, sandholtz2019measuring, hu2021bayesian}. These tools help bridge the gap between raw data and actionable insights, enabling a deeper understanding of one of basketball’s most critical aspects.

In professional basketball, understanding shooting patterns is essential for evaluating player performance and shaping team strategies. Shot charts, graphical depictions of shot locations on the court, are fundamental resources for analyzing these patterns. They provide a clear visualization of where players excel, how they distribute their attempts across the court, and their effectiveness in different regions, making them invaluable for assessing individual strengths and weaknesses. Beyond individual performance, shot charts inform coaching decisions and team strategies. Coaches use them to identify high-probability shooting zones, adjust offensive plays, and design defensive schemes to disrupt opponents’ preferred areas. For players, shot charts offer targeted feedback to refine shooting mechanics or expand scoring versatility. For the general manager of the team, they can target a similar shooting pattern with a lower salary. It provides a business strategy to the team owners especially for small market team. The evolution of basketball analytics, powered by player tracking data and advanced statistical methods, has elevated the importance of shot charts. Once simple visual aids, they now support sophisticated models that quantify shooting tendencies, evaluate defensive impact, and predict future performance. This transformation has made shot chart analysis a dynamic tool for uncovering patterns and driving data-informed decisions. However, analyzing shot charts is challenging due to the spatial and contextual complexity of shooting behavior. The shot distributions of players are shaped by factors such as the positioning of the court, the dynamics of the teammates, the dynamics of the teammates, and the defensive pressure. Addressing these nuances requires advanced statistical approaches that capture spatial correlations and variability, making shot chart analysis a key area of innovation in basketball analytics.

From a statistical perspective, shot chart data presents a unique challenge due to its inherently spatial nature. Shooting behavior is influenced by a combination of spatial factors, such as court positioning, defensive coverage, and team dynamics, which introduce complex correlations within the data. To address these complexities, researchers have developed a range of spatial statistical models. For example, \citet{reich2006spatial} proposed a multinomial logit regression framework with spatially varying coefficients to explore how contextual variables, such as game location, defensive intensity, and teammate interactions, impact shot-making probabilities in different court regions. More recently, spatial point process models have gained traction for analyzing shot chart data, reflecting the stochastic nature of shot locations \citep{miller2014factorized, jiao2019bayesian,yin2022bayesian, wong2023joint, qi2025aoas}. These models treat shot attempts as random spatial events and allow for more flexible representations of shooting behavior. By accounting for spatial randomness and incorporating external factors, these approaches have advanced our ability to quantify and predict shooting patterns. However, despite these advances, further innovation is needed to fully unravel the complexities of shooting data, particularly in terms of modeling player-specific tendencies and uncovering latent patterns that drive performance variability.

Another common starting point for this analysis is the creation of intensity surfaces \citep{yin2023analysis,hu2021bayesian}, which summarize shooting frequency or success rates across spatial regions of the court. Log Gaussian Cox Process \citep[LGCP, ][]{moller1998log} is used to model the spatial pattern of those shot attempts for each player. intensity surfaces provide a visual representation of a player's preferred shooting zones, revealing both strengths and tendencies. In this article, we aim to investigate the relationship between intensity surfaces and both the player's characteristics and the game contextual information. Unlike real-valued variables, the intensity surfaces exhibit clear spatial inner structures and, therefore, fall within the realm of functional data \citep[e.g.,][]{ramsay1991some,ramsay2005functional,wang2016functional}. Thus, we can formalize the problem as a function-on-scalar regression (FOSR) problem, where the responses are two-dimensional intensity surfaces, and the covariates are scalars. Despite the various models developed in FOSR, the nonstationarity of intensity patterns and the nonlinear regression relationship between intensity surfaces and scalar predictors pose new challenges for effectively modeling shot intensities and performing uncertainty quantification. In particular, existing research primarily focuses on functional linear models \citep{morris2006wavelet,rosen2009bayesian,morris2015functional,chen2016variable,kowal2020bayesian,ghosal2023shape}, which assume a linear relationship between the response function and the covariate vector.

In this article, we propose Adaptive Functional Bayesian Additive Regression Trees (AFBART) to address the aforementioned challenges in regression analysis with intensity surfaces, and develop a tailored Bayesian backfitting algorithm for posterior sampling. This approach is motivated by the success of Bayesian additive regression trees \citep[BART][]{chipman2010bart} and their recent extension to functional regression \citep[FBART,][]{cao2025functional}. Over the past decade, Bayesian tree-based models have gained significant popularity due to their flexible nonparametric nature, empirical success across various disciplines, and easy accessibility to uncertainty measures \citep{linero2018bayesian,hahn2020bayesian, hill2020bayesian,rovckova2020posterior,starling2020bart,luo2021bast,li2023adaptive}. In the sports analytics literature, decision-tree-based methods have been introduced in recent years \citep[e.g., ][]{huang2020regression, zuccolotto2023spatial} and have shown success in analyzing NBA games and player performance. However, their Bayesian counterparts like BART models, which offer a formal Bayesian framework for uncertainty quantification, have not been thoroughly investigated in sports analytics. Recently, \cite{yee2024scalable} introduced ridgeBART, an extension of BART, to incorporate smoothness with respect to targeted covariates. The model is applied to estimate the field goal percentage of NBA players, assuming smooth variation with respect to shot location. \cite{cao2025functional} proposed FBART for function-on-scalar regression problems with possible shape constraints using a B-spline representation, making it readily applicable for our current regression analysis on shot intensity surfaces.

AFBART enhances multivariate function representation by using an adaptive reduced-rank basis and facilitating highly nonlinear regression through a tree ensemble. Unlike FBART, which relies on pre-specified basis functions that are challenging to select for multidimensional shot intensity surfaces with evident spatial nonstationarity, AFBART employs adaptive basis functions derived directly from the data. This method builds on the foundation of nonlinear FOSR established by FBART and extends it by integrating Bayesian adaptive basis modeling, which aligns with techniques such as Bayesian functional principal component analysis \citep{greven2011longitudinal} and has been utilized in linear FOSR challenges \citep{kowal2020bayesian}. To our knowledge, AFBART is the inaugural fully nonparametric Bayesian approach in FOSR that adopts an adaptive basis representation. Furthermore, AFBART incorporates Bayesian additive regression trees to model the relationship between shot intensity functions and an extensive array of covariates dynamically. This comprehensive framework supports detailed analysis of model fitting outcomes, exploring shooting patterns of prominent players, assessing synthetic representatives, and conducting a variable importance analysis concerning player attributes and contextual game factors.

The rest of the paper is organized as follows. In Section \ref{sec:motivating_data}, we briefly discuss the player's shooting data from 2017–2018 National Basketball Association (NBA) regular season. In Section \ref{sec:method}, we introduce the model framework for Bayesian Adaptive Functional Additive Regression Trees model (AFBART). Details of the Bayesian inference procedure, including the MCMC algorithm and prior specifications, are presented in Section \ref{sec:algorithm}. A comprehensive simulation studies are conducted in Section \ref{sec:simu}. Real data applications of the proposed methods to NBA players data are provided in Section \ref{sec:real_data}. Section \ref{sec:discuss} concludes the paper with a discussion and future works.

\section{Motivating data}\label{sec:motivating_data}
Our analysis is based on data from the 2017–2018 National Basketball Association (NBA) regular season. The dataset includes a wide range of summary statistics, such as player position, age, team affiliation, minutes played (MP), player efficiency rating (PER), and field goal attempt rate. These metrics provide a comprehensive overview of player performance and usage during the season. A key component of the dataset is the shot intensity surfaces, which capture the spatial distribution of both made and missed field goals. These surfaces were obtained from \url{http://nbasavant.com/index.php}, a resource known for its detailed shot-tracking data. Additionally, player summary statistics were sourced from \url{https://www.basketball-reference.com}, a reputable platform for basketball analytics. To ensure the relevance and reliability of our analysis, we focused on players who attempted more than 400 field goals (FGA) during the season. This threshold helps filter out players with limited playing time or inconsistent roles, ensuring that the analysis reflects the performance of key contributors. Furthermore, we excluded players who began their careers in the 2017–2018 season, such as Lonzo Ball and Jayson Tatum, as their limited experience could skew the results. After applying these criteria, a total of 191 players were included in our study, providing a robust sample for analysis.

\begin{figure}[h]
    \centering
    \includegraphics[width=\linewidth]{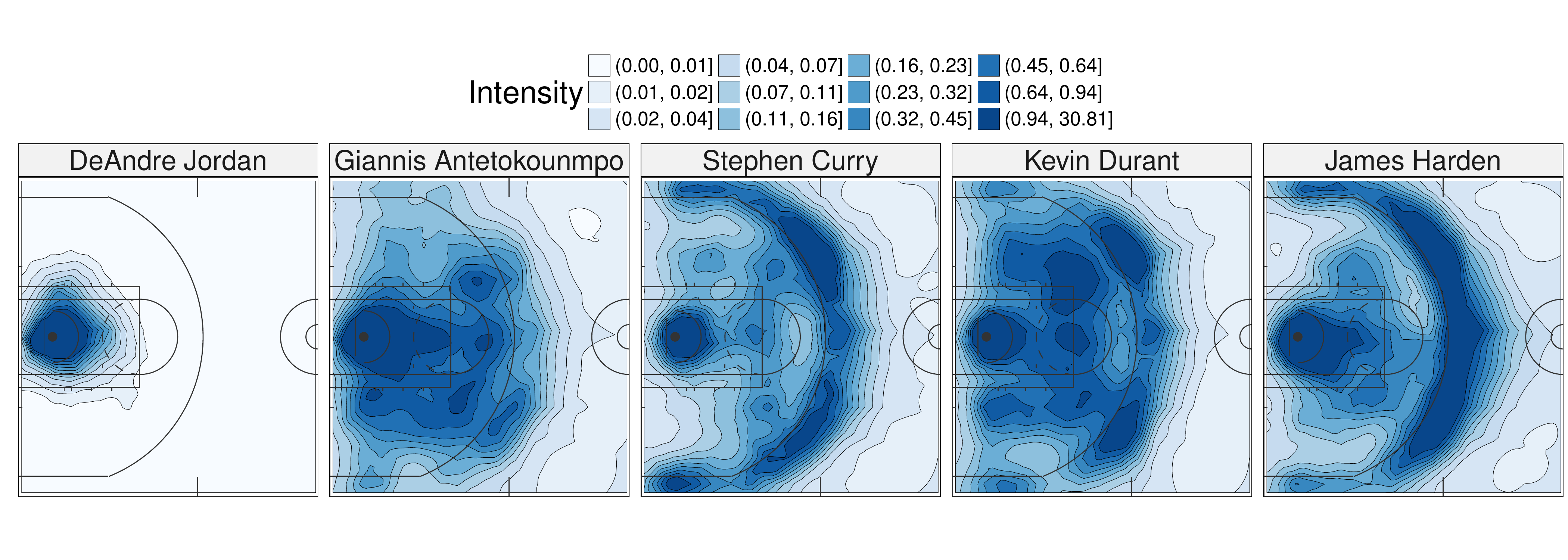}
    \caption{The shot intensity surfaces of five representative NBA players.}
    \label{fig:data:intensity}
\end{figure}

For each player $i \in \left\{1,\ldots, 191\right\}$, an intensity surface $\lambda_i(\cdot)$ is constructed to represent their shot locations on the court. This surface maps each point on the half-court domain $\sB$ to a real-valued intensity, reflecting the frequency and distribution of shot attempts. Additionally, a set of covariates $\vx_i \in \R^p$ is collected for each player, encompassing 24 summary statistics that capture various aspects of their performance and playing style. These statistics include metrics such as minutes played, player efficiency rating (PER), and field goal attempt rate, among others. The intensity surfaces $\lambda_i(\cdot)$ are discretized over a half-court domain $D \in [0, 72] \times [0, 74]$, divided into a grid of $72 \times 74$ boxes. This discretization allows for a detailed spatial analysis of shooting patterns. The intensity for each player is modeled using a Log-Gaussian Cox Process (LGCP), a statistical framework that accounts for spatial randomness and correlation in shot locations. For a deeper understanding of the LGCP methodology, refer to \citet{hu2021bayesian,yin2023analysis}. To demonstrate the application of this approach, we selected five players with distinct playing styles and preferred positions: DeAndre Jordan (center), Giannis Antetokounmpo (forward), Stephen Curry (guard), Kevin Durant (forward), and James Harden (guard). These players were chosen to highlight the diversity in shot selection and spatial patterns across different roles on the court. A subset of their summary statistics is presented in Table~\ref{tab:player_stat}, while their corresponding intensity surfaces are visualized in Figure~\ref{fig:data:intensity}. These visualizations provide insights into each player's shooting tendencies, such as preferred shot locations and areas of high activity.

In Table~\ref{tab:player_stat}, we present a detailed overview of player statistics, including each player's name, preferred position, minutes played, true shooting percentage (TS\%), and win shares (WS). True shooting percentage (TS\%) is a advanced metric that evaluates shooting efficiency by incorporating 2-point field goals, 3-point field goals, and free throws. Unlike traditional field goal percentages, which only consider one type of shot, TS\% offers a more holistic assessment of a player's scoring efficiency. For example, a player who excels at drawing fouls and making free throws may have a higher TS\% than a player who relies solely on 2-point field goals, even if their field goal percentages are similar. Win shares (WS) is another advanced statistic that quantifies a player's overall contribution to their team's success. It estimates the number of wins a player contributes based on their individual performance. This metric allocates credit for a team's wins to each player, considering factors such as scoring, rebounding, assists, and defense. For more information on how win shares are calculated, refer to \url{https://www.basketball-reference.com/about/ws.html}. During the 2017-2018 season, James Harden led all players with a WS of 15.4, the highest in the league. Giannis Antetokounmpo ranked sixth, and Kevin Durant ranked eighth, highlighting their significant contributions to their respective teams. In terms of shooting efficiency, a TS\% above 60\% is generally considered elite, reflecting exceptional scoring ability. The league average TS\% for the 2017-2018 season was 55.6\%, and all players listed in Table~\ref{tab:player_stat} exceeded this average, underscoring their status as some of the most efficient scorers in the league.

DeAndre Jordan, a traditional center, is known for his dominance in the paint and his lack of three-point attempts. His offensive game revolves around the restricted area, where he consistently scores points through dunks, layups, and putbacks. This focus on high-percentage shots contributes to his impressive field goal percentage, ranking him second in the league. Additionally, his physical presence and positioning make him a rebounding force, also placing him second in rebounds. Giannis Antetokounmpo, despite having the lowest true shooting percentage (TS\%) among the five players discussed, still maintains an elite TS\%, reflecting his efficiency as a scorer. His ranking of sixth in win shares (WS) highlights his overall impact on the game, which extends beyond scoring. Giannis excels in other areas, such as rebounds and blocks, ranking tenth in total rebounds and thirteenth in total blocks. These statistics underscore his versatility and defensive prowess. Stephen Curry and James Harden are both known for their high-volume three-point shooting. Their shot selection reflects a modern NBA approach, with most attempts coming from beyond the arc or in the paint. During the 2017-2018 season, Stephen Curry led the league in true shooting percentage (TS\%), showcasing his exceptional efficiency. James Harden, on the other hand, led in three-point attempts, points per game, and player efficiency rating (PER). Harden’s all-around contributions are further evidenced by his top-10 rankings in assists and steals, highlighting his role as both a scorer and playmaker. Kevin Durant, one of the league’s most versatile offensive players, ranks in the top 10 in TS\%, blocks, and points. Compared to Stephen Curry and James Harden, Durant’s three-point attempt rate is notably high, reflecting his ability to score from anywhere on the court. As one of the top offensive players in the league, Durant’s shooting attempts are distributed across all zones, making him a constant threat regardless of defensive coverage.

\begin{table}[h!]
\caption{A subset of summary statistics (Games: Games played or pitched, MP: Minutes Played, TS\%: True Shooting Percentage, 3PAr: 3-Point Attempt rate, WS: Win Shares). }
\centering
\begin{tabular}{lccccccc}
\toprule
\textbf{Player} & \textbf{Position} & \textbf{Age} & \textbf{Games} & \textbf{MP} & \textbf{TS\%} & \textbf{3PAr} & \textbf{WS} \\
\midrule
DeAndre Jordan        & Center  & 29 & 77 & 2424 & 0.648 & 0.000 & 9.4  \\
Giannis Antetokounmpo & Power Forward & 23 & 75 & 2756 & 0.598 & 0.100 & 11.9 \\
Stephen Curry         & Point Guard & 29 & 51 & 1631 & 0.675 & 0.580 & 9.1  \\
Kevin Durant          & Small Forward & 29 & 68 & 2325 & 0.640 & 0.338 & 10.4 \\
James Harden          & Shooting Guard & 28 & 72 & 2551 & 0.619 & 0.498 & 15.4 \\
\bottomrule
\end{tabular}

\label{tab:player_stat}
\end{table}

\section{Methodology}\label{sec:method}
We first introduce the notations used in this paper. Let the symbol $\|\cdot\|_q$ denote the $q$-norm of vectors and matrices, for $q \in [1, \infty]$. For a positive integer $j$, we use $[j]$ to denote the set of consecutive integers $\{1, 2, \ldots, j\}$. For a vector $\vb$, we use $\vb(i)$ to represent its $i$-th entry. For a matrix $\mA$, $\mA(i, j)$ denotes its $(i, j)$-th element. We use $\mb{0}$ to denote the zero vector and $\mathbf{I}_n$ to denote the identity matrix of size $n$. We use $\sN(\cdot, \cdot)$ to denote a (multivariate) normal distribution, and $\mathcal{N}(\cdot;\cdot, \cdot)$ to denote the corresponding density function. We use $p(\cdot)$, $\pi(\cdot)$, and $\Pi(\cdot)$ to denote the likelihood function, prior distribution, and posterior distribution, respectively. Given a set $A$, $\bbI_{A}(\cdot)$ denotes the indicator function on $A$.

\subsection{Regression between intensity surfaces and scalar covariates}

We formalize the regression analysis of intensity surfaces as a Function-on-Scalars Regression (FOSR) problem. For each player $i \in [n]$, we observe an intensity surface $\lambda_i(\cdot)$, which is a real function with domain $\sB\subset\R^2$ representing the half court, and the corresponding covariates $\vx_i \in \R^p$, summarizing the player's performance and game contextual information. We assume the following FOSR model to describe the association between intensity surface $\lambda_i(\cdot)$ and covariate $\vx_i$:
\begin{equation}\label{eq:modeling:fosr}
    Z_i(\cdot) = \log(\lambda_i(\cdot)) = \Xi_{\vx_i}(\cdot) +e_i,\quad i=1,2,\ldots,n,
\end{equation}
where $\Xi_{\vx}(\cdot) = \E(\log \lambda(\cdot)\mid \vx)$ is the true conditional mean function and $e_i$ is the independent Gaussian white noise process with variance $\sigma^2>0$.

To efficiently represent the mean functions, we further assume $\Xi_{\vx}(\cdot)$ exhibits basis expansion 
\begin{equation}\label{eq:FBART:mean}
    \Xi_{\vx}(\cdot) = \vf\trans(\cdot)\vh(\vx),
\end{equation}
where $\vf(\cdot) = (f_1(\cdot), \ldots, f_J(\cdot))\trans$ is a vector of basis functions on $\sB$, and $\vh: \R^p \to \R^J$ is a multivariate regression function whose values are the corresponding basis coefficients. In the literature, $\vf$ can also be referred to as loading functions, and $\vh$ as the factors. 

In this work, we consider both $\vf$ and $\vh$ unknown and are modeled with nonparameteric approaches. Overall, the proposed model ``Adaptive Functional Bayesian Additive Regression Trees (AFBART)" consists of an adaptive reduced-rank basis representation for functional responses, an additive-tree model for the basis coefficient functions, and a set a regularization prior. 

\subsection{Adaptive basis functions}\label{subsec:adaptivebasis}

For the basis functions $\vf$, a straightforward strategy is to consider them as fixed. Examples include wavelets \citep{morris2006wavelet}, B-splines \citep{cao2025functional}, and functional principal components (FPC) basis \citep{goldsmith2013corrected}. However, when the domain is multidimensional, it typically requires a higher basis dimension to capture the data variation effectively. Moreover, while deploying FPC basis functions could yield a good fit, the uncertainty in FPC decompositions is commonly ignored \citep{goldsmith2013corrected, kowal2020bayesian}.

Instead, we adopt a set of adaptive functional basis functions as described in \cite{kowal2020bayesian}. Let
\begin{equation}\label{eq:modeling:fk}
    f_j(\cdot) = \vb\trans(\cdot) \vpsi_j, \quad \forall j \in [J],
\end{equation}
where $\vb$ is a vector of $K$-dimensional orthonormal low-rank thin plate splines (LR-TPS), which are well-defined for $\sB \subset \R^2$, and $\vpsi_j \in \R^K$ is the basis coefficient vector. Typically, we select a moderately large size of $K$ to effectively model the $J$-dimensional basis functions $\vf(\cdot)$.

The prior distributions of the basis coefficient vectors $\boldsymbol{\psi}_j$'s are specified to encourage the smoothness of basis functions $\vf(\cdot)$. In particular, we follow \cite{kowal2020bayesian} and adopt mixture of normal prior distributions:
\begin{equation}\label{eq:model:prior:basis}
    \boldsymbol{\psi}_j\mid \lambda_j \sim \sN\left(\mb{0}, \lambda_{j}^{-1} \mOmega^{-1}\right), \quad\pi(\lambda_j) \propto 1,
\end{equation}
where $\mOmega \in \R^{K \times K}$ is a pre-specified roughness penalty matrix, and $\lambda_j \in \R^+$ is a smoothing parameter. The penalty matrix $\mOmega$ is the $L_2$ inner product matrix of the second derivatives of basis functions $b_1(\cdot), \ldots, b_K(\cdot)$, as discussed in \cite{wood2017generalized} and \cite{kowal2018dynamic}, which is closely related to regression splines with thin plate bases. For the smoothing parameters $\lambda_j$, we adopt a vague hyperprior $\pi(\lambda_j) \propto 1$, allowing for conjugate updates. Detailed information about the LR-TPS basis and the penalty matrix can be found in Supplementary Section~S.2.

By definition, the mean function $\Xi_\vx(\cdot)$ is specified as the inner product between $\vf(\cdot)$ and $\vh(\vx)$, making it impossible to identify both $\vf$ and $\vh$ simultaneously. To alleviate this identification issue and improve the convergence of the posterior sampling algorithm, we adopt a matrix orthonormality constraint on $\vf$ (equivalently on $\mPsi$) as introduced by \cite{kowal2020bayesian}, which is demonstrated in Section~\ref{sec:algorithm}. Under this constraint, the unknown basis functions $\vf(\cdot)$ and coefficient functions $\vh(\vx)$ are identifiable up to orthonormal linear transformations.

\subsection{Bayesian Functional regression trees for basis coefficient functions}\label{subsec:FBART}
 
To model the unknown function $\vh$, we adopt a sum-of-trees model, motivated by Functional Bayesian Additive Regression Trees \citep[FBART,][]{cao2025functional}, to capture the highly nonlinear relationship between $Z_i(\cdot)$ and $\vx_i$. The idea is to represent piece-wise constant regression relationships with additive regression trees in the spirit of boosting, and to use regularization priors to prevent overgrown trees that could lead to overfitting.

We first introduce the notations for regression trees. Let $\mT$ denote a binary decision tree with $L$ terminal nodes. Specifically, a binary decision tree $\mT$ can be represented by a binary tree topology and a set of splitting rules for the internal nodes. The splitting rules are binary splits of the form $\{\vx:\vx(j)\leq z\}$ versus $\{\vx:\vx(j)> z\}$, where $\vx(j)$ is the splitting variable with $j\in[p]$ and $z\in\R$ is the splitting value. The terminal nodes (leaves) of $\mT$ then yield a partition $\sD = \{D^1,\ldots,D^{L}\}$ of the covariate space, where each cell $D^l$ is a rectangular-shaped subset. Given a binary decision tree $\mT$ with $L$ leaf nodes, we refer to the following map $\vg(\cdot; \mT, \sM): \R^p \to \R^J$ as a multivariate regression tree:
\begin{equation*}
    \vg(\cdot; \mT, \sM) = \sum_{\ell=1}^L \vmu_{\ell} \mathbb{I}_{D^{\ell}}(\cdot),
\end{equation*}
where $\sM = \{\vmu_1, \ldots, \vmu_L\} \subseteq \R^J$ is referred to as the corresponding node parameters. 

In the spirit of boosting, we define the multivariate additive regression tree as follows. Let $\{\mT_t\}_{t=1}^T$ denote a collection of $T \geq 1$ binary decision trees. For each $\mT_t$ with $L_t$ leaf nodes, the induced partition is denoted by $\sD_t = \{D_t^{\ell}\}_{\ell=1}^{L_t}$. Let $\sM_t = \{\vmu_{t\ell}\}_{\ell=1}^{L_t} \subseteq \R^J$ be the node parameters associated with $\mT_t$. By writing $\bbT = \{\mT_t\}_{t=1}^T$ and $\bbM = \{\sM_t\}_{t=1}^T$, the multivariate additive regression trees is defined as:
\begin{equation}\label{eq:modeling:mBART}
    \vg(\cdot; \bbT, \bbM) = \sum_{t=1}^T \vg(\cdot; \mT_t, \sM_t) = \sum_{t=1}^T \sum_{\ell=1}^{L_t} \vmu_{t\ell} \mathbb{I}_{D_t^{\ell}}(\cdot).
\end{equation}

To avoid overfitting, we adopt the regularization prior for binary decision trees $\bbT$ and node parameters $\bbM$ following \cite{cao2025functional}. In particular, the prior takes the form:
\begin{equation*}
    \pi\big(\{\mT_t,\sM_t\}_{t=1}^T\big) = \prod_{t=1}^T\pi( \sM_t\mid\mT_t)\pi(\mT_t).
\end{equation*}
The prior distributions of $\sigma^2$ and $\sM_t$ are specified to be the respective conjugate priors:
\begin{equation*}
    \pi(\sigma^2) \sim \nu\lambda/\chi^2_\nu, \quad \pi\big(\sM_t \mid\mT_t \big)  = \prod_{\ell=1}^{L_t}\sN(\vmu_{t \ell};\vmu_\mu,\mV_\mu),
\end{equation*}
where $\chi^2_\nu$ stands for the Chi-square distribution with degrees of freedom $\nu$. The mean parameter $\vmu_\mu\in\R^J$, the covariance matrix $\mV_\mu\in \R^{J\times J}$, $\lambda\in \R^{+}$, and $\nu\in\mathbb{N}^+$, are hyperparameters. 

For the prior distributions of $\mT_t$'s, we employ the tree-generating prior in \cite{cao2025functional}, which is modified based on \cite{chipman2010bart}. In particular, a binary decision tree $\mT$ follows $\pi(\mT)$ if $\mT$ can be generated by the following process: for each node at depth $d\geq 0$, it is non-terminal with probability $p_{\text{split}}(d) = a\gamma^{d}$, where $a\in(0,1]$ and $\gamma\in(0,1)$ are hyperparameters; and the splitting rule associated with each internal node is uniformly sampled from all possible splitting rules. The selection of such prior is motivated by \cite{rovckova2019theory} to ensure that $\pi(\mT_t)$ exhibits certain tail behaviors. For details on the tree-generating process, refer to \cite{cao2025functional}.

\begin{figure}
    \centering
    \includegraphics[width=0.9\linewidth]{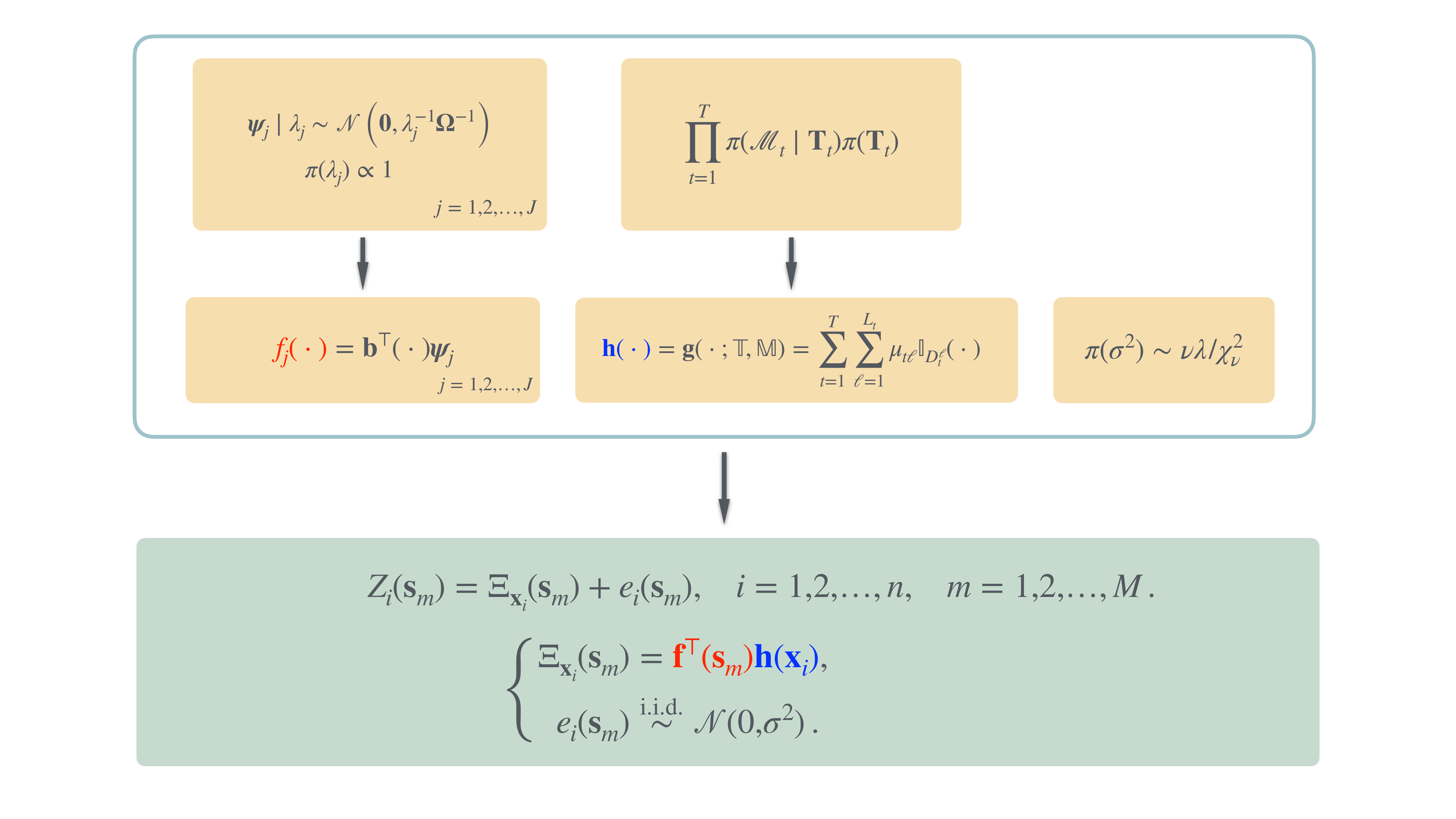}
    \caption{Model diagram for the proposed Adaptive Functional Bayesian Additive Regression Trees (AFBART).}
    \label{fig:model_diagram}
\end{figure}

Finally, the proposed Adaptive Functional Bayesian Additive Regression Trees model is obtained by combining the adaptive basis representation in Section~\ref{subsec:adaptivebasis} and tree-based modeling in Section~\ref{subsec:FBART}. Suppose the intensity surfaces $\lambda_i(\cdot)$ are observed at a set of spatial points $\mS = \{\vs_1, \ldots, \vs_M\} \subset \sB$ of size $M$. Define $\vz_i = (Z_i(\vs_1), \ldots, Z_i(\vs_M))\trans \in \R^M$, $\mF = (f_j(\vs_m))_{(m,j)} \in \R^{M \times J}$, $\vf_j = (f_j(\vs_1), \ldots, f_j(\vs_M))\trans \in \R^M$, $\mB = (b_k(\vs_m))_{(m,k)} \in \R^{M \times K}$, and $\mPsi = (\vpsi_1,\ldots, \vpsi_J)\in\R^{K\times J}$. By combining Equation~\eqref{eq:modeling:fosr}, Equation~\eqref{eq:modeling:fk}, and Equation~\eqref{eq:modeling:mBART}, we have for $i\in[n]$,
\begin{equation}\label{eq:modeling:AFAR}
    \vz_i = (\mB\mPsi) \vg(\vx_i; \bbT, \bbM) + \ve_i, \quad \ve_i \sim \sN(\mb{0}, \sigma^2 \mI_M),
\end{equation}
where the model parameters are $\{\mPsi, \bbT,\bbM, \sigma^2\}$. The model framework and prior specifications are summarized in Figure~\ref{fig:model_diagram}.

\section{Posterior sampling and prior specification}\label{sec:algorithm}

To sample from the posterior distribution of the AFBART model parameters, which are of considerable size, we develop an efficient Markov chain Monte Carlo (MCMC) algorithm. Specifically, our algorithm is a Gibbs sampler with Metropolis-Hastings steps for the discrete binary decision trees. We summarize the sampling process in Algorithm~\ref{alg:AFBART} below.

\begin{algorithm}[h]
\caption{Sampling from the AFBART posterior distribution}\label{alg:AFBART}
\KwIn{ Data $\{\vx_i,\vz_i\}_{i=1}^n$, number of iterations $n_{\mathrm{mcmc}}$, initial values of $\mPsi$,$\bbT$,$\bbM$, $\sigma^{2}$, and $\{\lambda_j\}_{j=1}^J$.}
\For{$i_{\mathrm{mcmc}}\in[n_{\mathrm{mcmc}}]$}{
    1. Update $\bbT$ and $\bbM$: \For{$t\gets 1$ \KwTo $T$}{
        (i). Update $\mT_t$: Sample a new $\mT^*_t$ from the proposal distribution $q(\mT_t, \mT^*_t)$, and accept the new sample and update $\mT_t= \mT^*_t$ with probability
                \begin{equation*}
                    \alpha\left(\mT_t, \mT^*_t\right)=\min \left\{\frac{q\left(\mT_t^*, \mT_t\right)}{q\left(\mT_t, \mT_t^*\right)} \frac{p\big(\vz_1,\ldots,\vz_n\mid \mPsi, \mT_t^*,\bbT_{(t)},\bbM_{(t)},\sigma^2\big) \pi\left(\mT_t^*\right)}{p\big(\vz_1,\ldots,\vz_n\mid \mPsi, \mT_t,\bbT_{(k)},\bbM_{(k)},\sigma^2\big) \pi\left(\mT_t\right)}, \quad 1\right\}.
                \end{equation*}
            The marginal likelihood $p(\vz_1,\ldots,\vz_n \mid \mPsi, \mT_t, \bbT_{(t)}, \bbM_{(t)}, \sigma^2)$ over $\sM_t$  takes the form
            \begin{equation*}
             \prod_{\ell=1}^{L_t} \frac{(2\pi\sigma^2)^{-\frac{n_{t\ell}}{2}} |\mV_\mu|^{-\frac{1}{2}}}{|\mV_{\text{post}}^{t\ell}|^{-\frac{1}{2}}}\exp \left[ \frac{1}{2} (\vmu_{\text{post}}^{t\ell})\trans (\mV_{\text{post}}^{t\ell})^{-1} \vmu_{\text{post}}^{t\ell}
            - \frac{1}{2\sigma^2} \sum_{i : \vx_i \in D_t^\ell} \vr_i\trans \vr_i - \frac{1}{2} \vmu_\mu\trans \mV_\mu^{-1} \vmu_\mu \right],
            \end{equation*}
            where
            $n_{t\ell} = \big| \{i : \vx_i \in D_t^\ell\} \big|$ is the number of observations whose covariate vector falls into the subregion $D_t^\ell$,  $\vr_i =  \vz_i - \mB\mPsi\vg(\vx_i; \bbT_{(t)}, \bbM_{(t)})$ is the partial residual, and 
            \begin{equation*}
                \mV_{\text{post}}^{t\ell} = \left[\mV_\mu^{-1} + \frac{n_{t\ell}}{\sigma^2}\mPsi\trans\mB\trans\mB\mPsi\right]^{-1}, \quad \vmu_{\text{post}}^{t\ell} = \mV_{\text{post}}^{t\ell}\left[\frac{1}{\sigma^2}\sum_{i:\vx_i \in D_t^\ell}(\mB\mPsi)\trans\vr_i + \mV_\mu^{-1}\vmu_\mu\right],
            \end{equation*}
            
        (ii). Update $\sM_t$: For each $\ell\in[L_t]$, sample $\vmu_{t\ell}$ from $\sN(\vmu_{\text{post}}^{t\ell}, \mV_{\text{post}}^{t\ell})$.
    }
    2. Update $\sigma^2$ by sampling from inverse-gamma distribution
    \begin{equation*}
        \sigma^2 \sim \mathrm{InvGamma}\left( \frac{\nu + nM}{2}, \frac{\lambda\nu + \sum_{i=1}^n \|\vz_i - \mB\mPsi\vg_i\|_2^2}{2} \right).
    \end{equation*}
    
    3. Update $\mPsi$: \For{$j\gets 1$ \KwTo $J$}{
            Sample $\vpsi_j$ from $\sN(\vmu_{\vpsi_j}, \mSigma_{\vpsi_j})$ conditional on $\mB\vpsi_j\perp \mB\vpsi_{j'}$ for all $j'\neq j$, where
            \begin{equation*}
            \begin{aligned}
                \mSigma_{\vpsi_j} &= \left( \sigma^{-2} \mB\trans\mB \sum_{i=1}^N \vg_i^2(j) + \lambda_{j}\mOmega \right)^{-1},\\
                \vmu_{\vpsi_j} &= \sigma^{-2} \mSigma_{\vpsi_j} \mB\trans \left[ \sum_{i=1}^N \vg_i(j) \left( \vz_i - \sum_{j' \neq j} \mB\vpsi_{j'}\vg_i(j') \right) \right].
            \end{aligned}
            \end{equation*}

            Normalize $\vpsi_j \gets \vpsi_j / \|\mB\vpsi_j\|_2$.
        }
    4.  Update $\{\lambda_j\}_{j=1}^J$: \For{$j\gets 1$ \KwTo $J$}{
            Sample $\lambda_j$ from $\mathrm{Gamma}\left(\frac{K}{2} + 1, \frac{\vpsi_j\trans\mOmega\vpsi_j}{2}\right)$.
        }
    }
\Return{Samples of parameters $\Big\{\mPsi, \bbT,\bbM, \sigma^2,\{\lambda_j\}_{j=1}^J\Big\}$.}
\end{algorithm}

First, given observed data $\{\vz_i\}_{i=1}^n$, the likelihood function can be obtained from model~\eqref{eq:modeling:AFAR}:
\begin{equation*}
    p(\vz_1,\ldots,\vz_n \mid \mPsi, \bbT, \bbM, \sigma^2) \propto (\sigma^2)^{-nM/2} \exp \left( -\frac{1}{2\sigma^2} \sum_{i=1}^n \|\vz_i - \mB\mPsi\vg_i\|_2^2 \right),
\end{equation*}
where we write $\vg_i := \vg(\vx_i;\bbT,\bbM)$ for simplicity.

In step 3 of Algorithm~\ref{alg:AFBART}, we update the coefficient vectors in $\mPsi = (\vpsi_1, \ldots, \vpsi_J)$ by sampling from their full conditional distributions. In particular, note that
\begin{equation*}
    (\mB\mPsi) \vg(\vx_i; \bbT, \bbM) = \left( \vg\trans(\vx_i; \bbT, \bbM) \otimes \mB \right) \mathrm{Vec}(\mPsi),
\end{equation*}
where $\otimes$ denote the kronecker product. According to our prior specification, the conditional prior of $\mathrm{Vec}(\mPsi)$ given $\mL$ follows $\sN(\mb{0}, \mL^{-1} \otimes \mOmega^{-1})$, where $\mL = \mathrm{diag}(\lambda_1, \ldots, \lambda_J)$. Following standard conjugacy results, we know that the full conditional distribution of $\mPsi$  is available in closed form, which follows a multivariate normal distribution. In Algorithm~\ref{alg:AFBART}, we sequentially update each $\vpsi_j$ from its respective full conditional distribution instead, to avoid directly sampling a relatively high-dimensional normal vector $\mathrm{Vec}(\mPsi)$ of dimension $KJ$. Furthermore, to enforce model identifiability, each $\vpsi_j$ is sampled from a constrained normal distribution with linear constraints $\mB\vpsi_j\perp \mB\vpsi_{j'}$ for all $j'\neq j$, and then normalized to ensure $\|\mB\vpsi_j\|_2 = 1$. As a result, the value matrix $\mF$ of basis functions $f_j(\cdot)$ is orthonormal in the sense that $\mF\trans\mF = (\mB\mPsi)\trans\mB\mPsi = \mI_J$.

For the binary decision trees $\mT_t$, which are highly discrete and structured, we update them with Metropolis-Hastings steps. Specifically, we sample a new tree $\mT^*_t$ from the proposal distribution $q(\mT_t, \mT^*_t)$ and accept the new sample with probability $\alpha(\mT_t, \mT^*_t)$. We adopt the proposal distribution described in \cite{cao2025functional}, and calculate the marginal likelihood over $\sM_t$ as well as the acceptance probability $\alpha(\mT_t, \mT^*_t)$  based on Lemma 1 in \cite{cao2025functional}. For the node parameters $\sM_t$, they follow normal full conditional distributions. The above updates for $\mT_t$ and $\sM_t$ are summarized in step~1 and step~2 of Algorithm~\ref{alg:AFBART}, respectively.

Finally, the full conditional distributions of the coefficient vectors $\lambda_j$ and the variance parameter $\sigma^2$ can be obtained using standard conjugacy arguments, as shown in step~2 and step~4 of Algorithm~\ref{alg:AFBART}, respectively.

To apply Algorithm~\ref{alg:AFBART}, we need to specify the hyperparameters in the model. We will use the following default settings, which can be adjusted via cross-validation in practice. For the prior distribution of $\mT_t$ and $\sM_t$, we adopt the data-informed prior described in \cite{cao2025functional} with 
 $f_j(\cdot)$ chosen as the first $J$ functional principal components based on the observed curves $\{Z_i\}$. For the prior of $\sigma^2$, we suggest setting $\nu = 3$ and choosing $\lambda$ such that $\pi(\sigma^2 < \hat{\sigma}^2) = 0.9$, where $\hat{\sigma}^2$ is the sample variance of all observed values $\{Z_i(\vs_{m})\}_{i\in[n],m\in[M]}$. We set the number of trees $T$ to be 50, the number of basis functions $J$ to be 20, and the number of thin plate bases $K$ from $\{50, 70, 100\}$. By default, the knots of thin plate basis functions are chosen using a space-filling algorithm. Additional implementation details are deferred to Supplementary Section~S.2.

\section{Simulation}\label{sec:simu}

\subsection{Simulation setup}
In this section, we evaluate the performance of AFBART using numerical experiments and compare it with several competitive models. We consider the FOSR model given by Equation~\eqref{eq:modeling:fosr} with $p=3$ covariates, where the response functions are defined on $[0,1]^2$. 

To generate the basis functions $\vf(\cdot)$ and coefficient functions $\vh(\cdot)$, we consider the following three cases. In Case 1, the $J=5$ basis functions $\vf(\cdot)$ are be generated as random linear combinations of $K=40$ thin plate basis functions on $[0,1]^2$:
\begin{equation*}
    f_j(\cdot)  = \vb\trans(\cdot)\vphi_j, \quad \vphi_j \overset{\mathrm{i.i.d.}}\sim \sN(0, \mI_K).
\end{equation*}
The coefficient functions $\vh(\cdot)$ are chosen to be piece-wise constant functions
\begin{equation*}
    h_{j}(\vx) = \mathbb{I}_{[0,0.5)}(\vx(1)) + \mathbb{I}_{[0,0.5)}(\vx(2))\big(1 + \mathbb{I}_{[0,0.5)}(\vx(3))\big),\quad j = 1,2,\ldots,J.
\end{equation*}
In this case, the coefficient functions can be represented using a single regression tree, and the basis functions can be modeled exactly by thin plate basis functions. 

In Case 2, the basis functions $\vf(\cdot)$ are randomly generated from a Gaussian process:
\begin{equation*}
    f_j(\cdot) \overset{\mathrm{i.i.d.}}\sim \mathcal{GP}\Big(0, \kappa(\cdot,\cdot)\Big),\quad j = 1,2,\ldots,J,
\end{equation*}
where $\kappa(\vs, \vs') = \exp(-\frac{9}{2}\|\vs-\vs'\|_2^2) + 10^{-6}\mathbf{1}_{\{\vs = \vs'\}}$ is an exponential covariance kernel with a nugget effect. The coefficient functions are chosen to be smooth nonlinear functions:
\begin{equation*}
    h_{j}(\vx) = 5\sin\Big(\pi \vx(1)\vx(2) \Big) + 10\Big(\frac{\vx(3)(j+2)-3}{6} \Big)^2,\quad j = 1,2,\ldots,J.
\end{equation*}
In this case, the proposed model is misspecified, meaning that the data-generating process can only be approximated by the model.

Finally, in Case 3, we follow the same settings as in Case 2, but specify the basis functions $\vf(\cdot)$ as the $J=5$ selected shot intensity surfaces depicted in Figure~\ref{fig:data:intensity}. In this case, AFBART is still misspecified, and the data is designed to mimic the real pattern of shot intensity surfaces.

\subsection{Competitive models and evaluation criteria}
We compare the proposed AFBART with several state-of-the-art competitive methods, including FBART \citep{cao2025functional}, the Bayesian FOSR method \citep[BFOSR,][]{kowal2020bayesian}, and the local linear regression methods with functional responses \citep[LLR, e.g.,][]{petersen2019frechet,fan2022conditional}. For AFBART, we set $J=5$ and $K = 40$. For the basis functions in FBART, we adopt $J=5$ thin plate spline basis functions. The hyperparameters in the above approaches, if not specified, are chosen according to their respective default settings. For all the additive tree models, we set the number of trees to be $T=50$. The MCMC algorithms are run for $2,000$ iterations for AFBART and FBART, and $4,000$ iterations for BFOSR. For all Bayesian methods, the final $400$ iterations are retained as posterior samples. For LLR, we use the normal kernel function, with the bandwidth chosen to minimize the in-sample root-mean-squared error. 

The prediction performance of different methods is quantified by three metrics. In particular, for each simulation setup, we independently generate $n^* = 200$ test data following the respective data-generating process. The first metric is the root mean squared prediction errors (RMSPE) defined as 
\begin{equation*}
\text{RMSPE} = \sqrt{\frac{1}{Mn^*}\sum_{i=1}^{n^*}\sum_{m=1}^{M}|\hat\Xi_{\vx^*_i}(\vs_m)-\Xi_{\vx^*_i}(\vs_m)|^2}, 
\end{equation*}
where $\vx^*_i$ are the covariate vectors from the test data, $\Xi_{\vx}(\vs)$ stands for the true function value, and $\hat\Xi_{\vx}(\vs)$ stands for the corresponding estimate. For Bayesian approaches, we use the posterior mean for point estimation. In addition, we calculate the pointwise posterior $95\%$ credible interval for uncertainty quantification. The accuracy of the credible interval is evaluated via the mean negatively oriented interval score \citep[MIS,][]{gneiting2007strictly}, defined as 
\begin{equation*}
\text{MIS} = \frac{1}{Mn^*}\sum_{i=1}^{n^*}\sum_{m=1}^{M}\Big[\hat U_{im}-\hat L_{im} + \frac{2}{5\%}\inf_{\eta\in [\hat L_{im},\hat U_{im}]} |\Xi_{\vx^*_i}(\vs_m)- \eta|\Big], 
\end{equation*}
where $\hat U_{im}$ and $\hat L_{im}$ are the $97.5\%$-quantile and $2.5\%$-quantile of the posterior samples of $\hat\Xi_{\vx^*_i}(\vs_m)$, respectively; note that for the frequentist approach LLR, MIS is calculated by setting $\hat U_{ij} = \hat L_{ij} = \hat\Xi_{\vx^*_i}(\vs_m)$. Last, we use the mean continuous ranked probability score \citep[MCRPS,][]{gneiting2007strictly} metric to evaluate the performance of probabilistic prediction. For all three metrics, a lower value indicates a better performance.

\subsection{Simulation results}

We apply the proposed AFBART method along with three competitive methods to simulated datasets. In particular, for each simulation case, we consider two different noise levels, $\sigma \in \{0.01, 0.1\}$, resulting in a total of six simulation settings. Under each setting, we independently generate $20$ replicated simulated datasets.

In Table~\ref{table:simu:main}, we report the average RMSPE, MIS, and MCRPS over $20$ simulation replicates. For each metric and simulation setting, the lowest metric value among the four methods is highlighted in \textbf{bold}. The results show that the proposed AFBART consistently outperforms the other methods across all three metrics, in both correctly specified and misspecified cases. Specifically, AFBART provides the most accurate point predictions, as indicated by the lowest RMSPE, while also delivering the highest-quality uncertainty quantification, as evidenced by the lowest MIS and MCRPS. Compared to FBART, AFBART leverages adaptive random basis functions to model functional responses, allowing it to dynamically search for the most representative basis functions. This advantage becomes particularly beneficial when dealing functional data with multidimensional domain, and there is insufficient prior knowledge to provide FBART with suitable low-rank basis functions. 

\begin{figure}[h]
    \centering
    \includegraphics[width=1\linewidth]{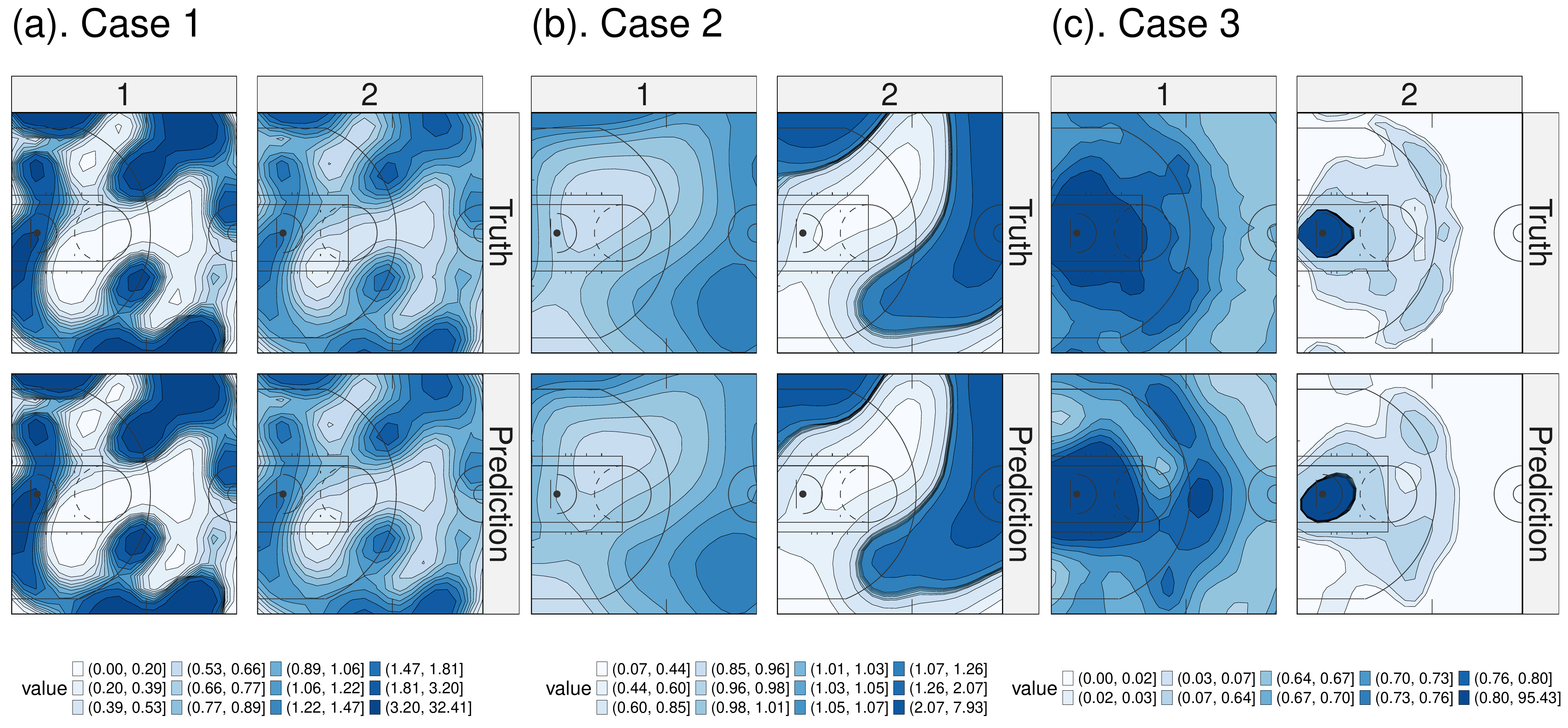}
    \caption{Out-of-sample predictions of AFBART across three simulation cases: Comparing the true function values with their respective posterior mean estimates.}
    \label{fig:simu:intensity}
\end{figure}

We further present the predicted functions of AFBART in Figure~\ref{fig:simu:intensity} with $\sigma = 0.01$. Under each simulation case, we select two covariate vectors $\vx_i^*$ from the test data and compare the exponential of the true response functions $\Xi_{\vx_i^{*}}(\cdot)$, denoted as ``Truth," with their corresponding AFBART posterior mean predictions, denoted as ``Prediction." Overall, the predictions closely align with the true function values, particularly in Case 1 and Case 2, where the true response functions are smooth.  In Case 3, where the response functions are designed to mimic the shot intensity surfaces of real NBA players and exhibit less smoothness, AFBART still produces predictions that remain close to the true values, while demonstrating a clear smoothing effect. This behavior can be attributed to the smoothness-encouraging prior in \eqref{eq:model:prior:basis} of the basis functions.

\begin{table}[h]
\centering
\caption{Comparison of AFBART, FBART, BFOSR, and LLR based on average RMSPE, MIS, and MCRPS over $20$ simulation replicates: Analysis across three simulation cases and two noise levels.}
\label{table:simu:main}
\centering
\begin{tabular}[t]{llrrrr}
\toprule
\multicolumn{2}{c}{ } & \multicolumn{4}{c}{Methods} \\
\cmidrule(l{3pt}r{3pt}){3-6}
\textbf{Setting} & \textbf{Metric} & \textbf{AFBART} & \textbf{FBART} & \textbf{BFOSR} & \textbf{LLR}\\
\midrule
\cellcolor{gray!10}{} & \cellcolor{gray!10}{RMSPE} & \cellcolor{gray!10}{\textbf{0.07}} & \cellcolor{gray!10}{0.68} & \cellcolor{gray!10}{13.38} & \cellcolor{gray!10}{20.31}\\
Case 1, $\sigma = 0.01$ & MIS & \textbf{0.37} & 18.47 & 6.76 & 76.04\\
\cellcolor{gray!10}{} & \cellcolor{gray!10}{MCRPS} & \cellcolor{gray!10}{\textbf{0.03}} & \cellcolor{gray!10}{0.36} & \cellcolor{gray!10}{0.15} & \cellcolor{gray!10}{1.90}\\
\cline{2-6}
 & RMSPE & \textbf{0.06} & 0.68 & 0.23 & 0.75\\
\cellcolor{gray!10}{Case 1, $\sigma = 0.1$ } & \cellcolor{gray!10}{MIS} & \cellcolor{gray!10}{\textbf{0.29}} & \cellcolor{gray!10}{18.47} & \cellcolor{gray!10}{2.68} & \cellcolor{gray!10}{19.13}\\
 & MCRPS & \textbf{0.06} & 0.36 & 0.14 & 0.45\\
 
\hline

 \cellcolor{gray!10}{} & \cellcolor{gray!10}{RMSPE} & \cellcolor{gray!10}{\textbf{0.04}} & \cellcolor{gray!10}{0.27} & \cellcolor{gray!10}{0.14} & \cellcolor{gray!10}{0.47}\\
Case 2, $\sigma = 0.01$ & MIS & \textbf{0.29} & 6.38 & 1.15 & 11.72\\
\cellcolor{gray!10}{} & \cellcolor{gray!10}{MCRPS} & \cellcolor{gray!10}{\textbf{0.02}} & \cellcolor{gray!10}{0.14} & \cellcolor{gray!10}{0.08} & \cellcolor{gray!10}{0.29}\\
\cline{2-6}
 & RMSPE & \textbf{0.04} & 0.27 & 0.14 & 0.39\\
\cellcolor{gray!10}{Case 2, $\sigma = 0.1$} & \cellcolor{gray!10}{MIS} & \cellcolor{gray!10}{\textbf{0.28}} & \cellcolor{gray!10}{6.38} & \cellcolor{gray!10}{1.14} & \cellcolor{gray!10}{10.35}\\
 & MCRPS & \textbf{0.06} & 0.15 & 0.10 & 0.24\\

 \hline

 \cellcolor{gray!10}{} & \cellcolor{gray!10}{RMSPE} & \cellcolor{gray!10}{\textbf{0.34}} & \cellcolor{gray!10}{0.69} & \cellcolor{gray!10}{0.90} & \cellcolor{gray!10}{3.02}\\
Case 3, $\sigma = 0.01$ & MIS & \textbf{3.58} & 10.13 & 7.95 & 80.92\\
\cellcolor{gray!10}{} & \cellcolor{gray!10}{MCRPS} & \cellcolor{gray!10}{\textbf{0.18}} & \cellcolor{gray!10}{0.36} & \cellcolor{gray!10}{0.53} & \cellcolor{gray!10}{2.02}\\
\cline{2-6}
 & RMSPE & \textbf{0.34} & 0.69 & 0.90 & 2.84\\
\cellcolor{gray!10}{Case 3, $\sigma = 0.1$} & \cellcolor{gray!10}{MIS} & \cellcolor{gray!10}{\textbf{3.54}} & \cellcolor{gray!10}{9.99} & \cellcolor{gray!10}{7.94} & \cellcolor{gray!10}{75.01}\\
 & MCRPS &\textbf{0.19} & 0.36 & 0.54 & 1.83\\
\bottomrule
\end{tabular}
\end{table}

\section{Deciphering field goal attempts habits of NBA players}\label{sec:real_data}
\begin{figure}[H]
    \centering
    \includegraphics[width=0.9\linewidth]{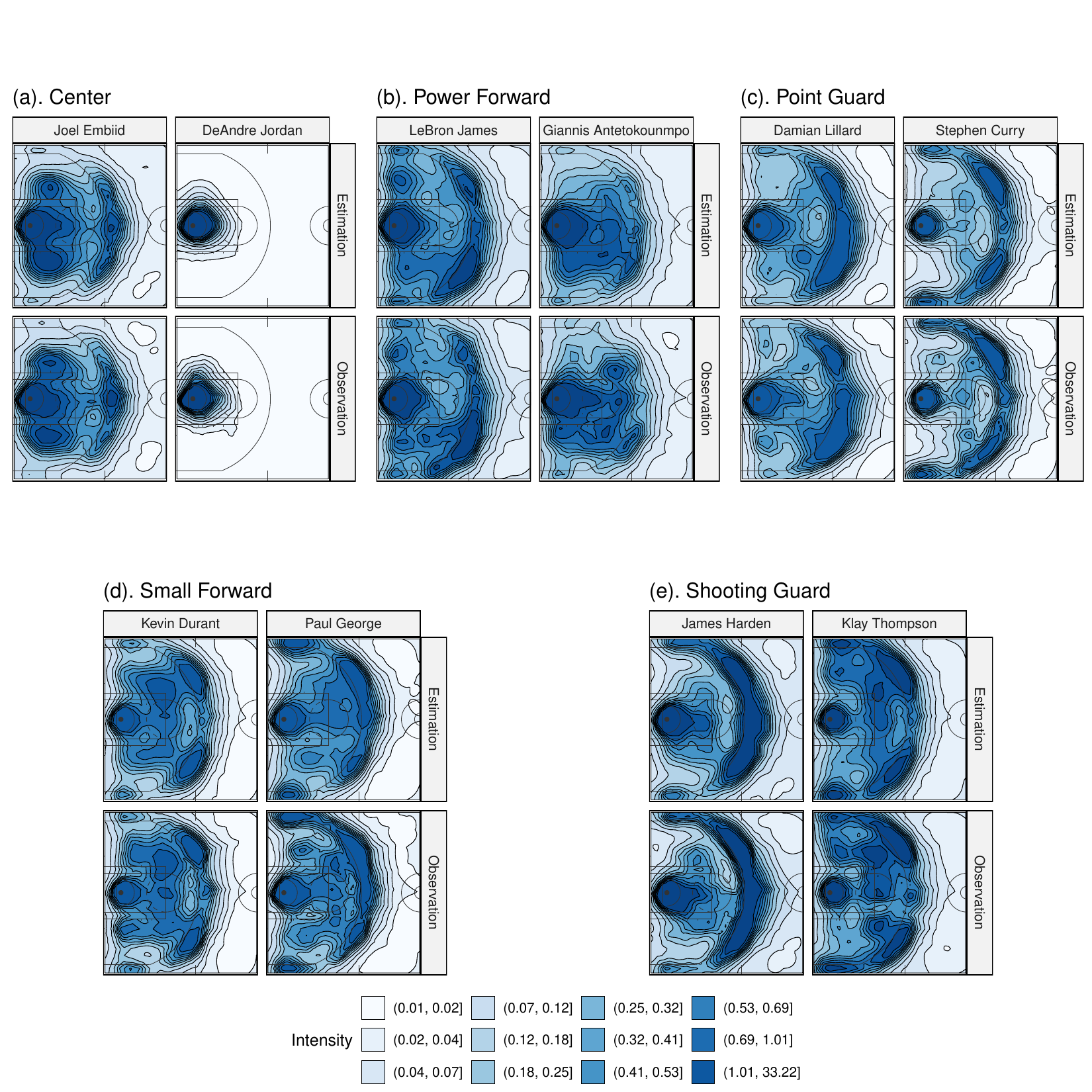}
    \caption{Shot intensity surfaces for selected players from different positions and their respective AFBART posterior mean estimates.}
    \label{fig:intensity_surface}
\end{figure}

We applied the AFBART approach to analyze the shot data for NBA players during the 2017-2018 regular season. For the AFBART approach, we set $T = 50$, $J = 20$, and $K = 70$. We ran the MCMC for $48,000$ iterations, discarding the first $40,000$ as burn-in. The remaining samples were retained as posterior samples with a thinning interval of $20$. We check it is sufficient for the chain to converge and stabilize. Figure~\ref{fig:intensity_surface} presents the posterior mean estimated intensity surfaces of ten representative players and compares them with the observed surfaces. We selected two players for each position. The selected players are Joel Embiid and DeAndre Jordan (Centers) in panel (a); Giannis Antetokounmpo and LeBron James (Power Forwards) in panel (b); Stephen Curry and Damian Lillard (Point Guards) in panel (c); Kevin Durant and Paul George (Small Forwards) in panel (d); and James Harden and Klay Thompson (Shooting Guards) in panel (e). We find that the AFBART consistently yields intensity surfaces that closely match the observed ones across all positions. 

We selected two different types of players for the center position. One is the traditional center, exemplified by players like DeAndre Jordan, who primarily focuses on the restricted area. These centers score most of their points through dunks, layups, and putbacks, dominating the paint. The other type is represented by players like Joel Embiid, who not only scores around the restricted area but also has a versatile offensive game. They are capable of making perimeter shots and/or three-pointers, adding more depth to their scoring ability and allowing them to stretch the floor. The power forward typically takes shots around the perimeter, with only a few attempts extending beyond the arc. While their primary role often involves mid-range shooting, many modern power forwards have also developed the ability to stretch the floor with occasional three-point attempts. This shift in the position has become more prominent as the game has evolved, with power forwards becoming more versatile and dynamic on offense. As a result, these players now contribute to both scoring inside and outside, creating mismatches with defenders and increasing their impact on the game. LeBron James is special case for power forward. As you may know, he can play at both power forward and small forward.

Small forwards like Kevin Durant who is known for their ability to score from nearly anywhere on the court. They are versatile offensive players, capable of shooting from mid-range, beyond the arc, and attacking the basket. In addition to their scoring skills, small forwards are also known for their physicality, often engaging in strong physical confrontations with defenders. Their combination of size, athleticism, and skill makes them a constant threat both offensively and defensively, allowing them to dominate in various aspects of the game. We selected Stephen Curry and Damian Lillard for the point guard position, both of whom rely heavily on high-volume three-point shooting. Their ability to stretch the floor and score from long range makes them constant offensive threats, and they also create opportunities for their teammates with their shooting gravity and playmaking skills. Compared to Stephen Curry and Damian Lillard, James Harden and Klay Thompson also extend their offensive threat to mid-range shooting. While Curry and Lillard are known for their deep three-pointers, Harden and Thompson are more versatile in their scoring, frequently hitting mid-range jumpers in addition to their long-range shots. They excel at catch-and-shoot situations, benefiting from quick ball movement and fast transitions. This ability to score both from beyond the arc and in the mid-range allows them to keep defenders off balance, making them dangerous on offense in various situations.

\begin{figure}[H]
    \centering
    \includegraphics[width=0.95\linewidth]{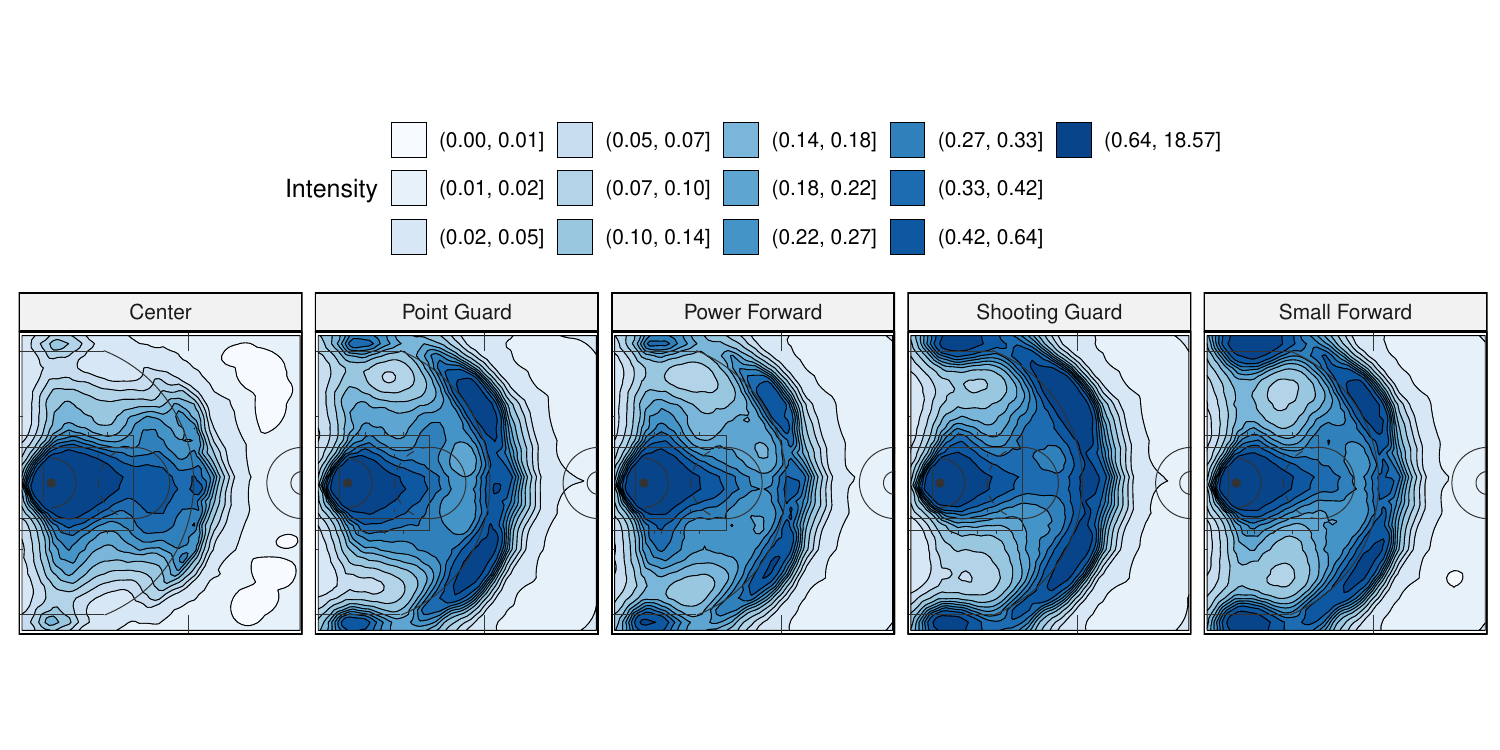}
    \caption{Estimated intensity surfaces for synthetic players with average covariate values for each position.}
    \label{fig:intensity_surface:positionavg}
\end{figure}

\begin{figure}[H]
    \centering
    \includegraphics[width=0.95\linewidth]{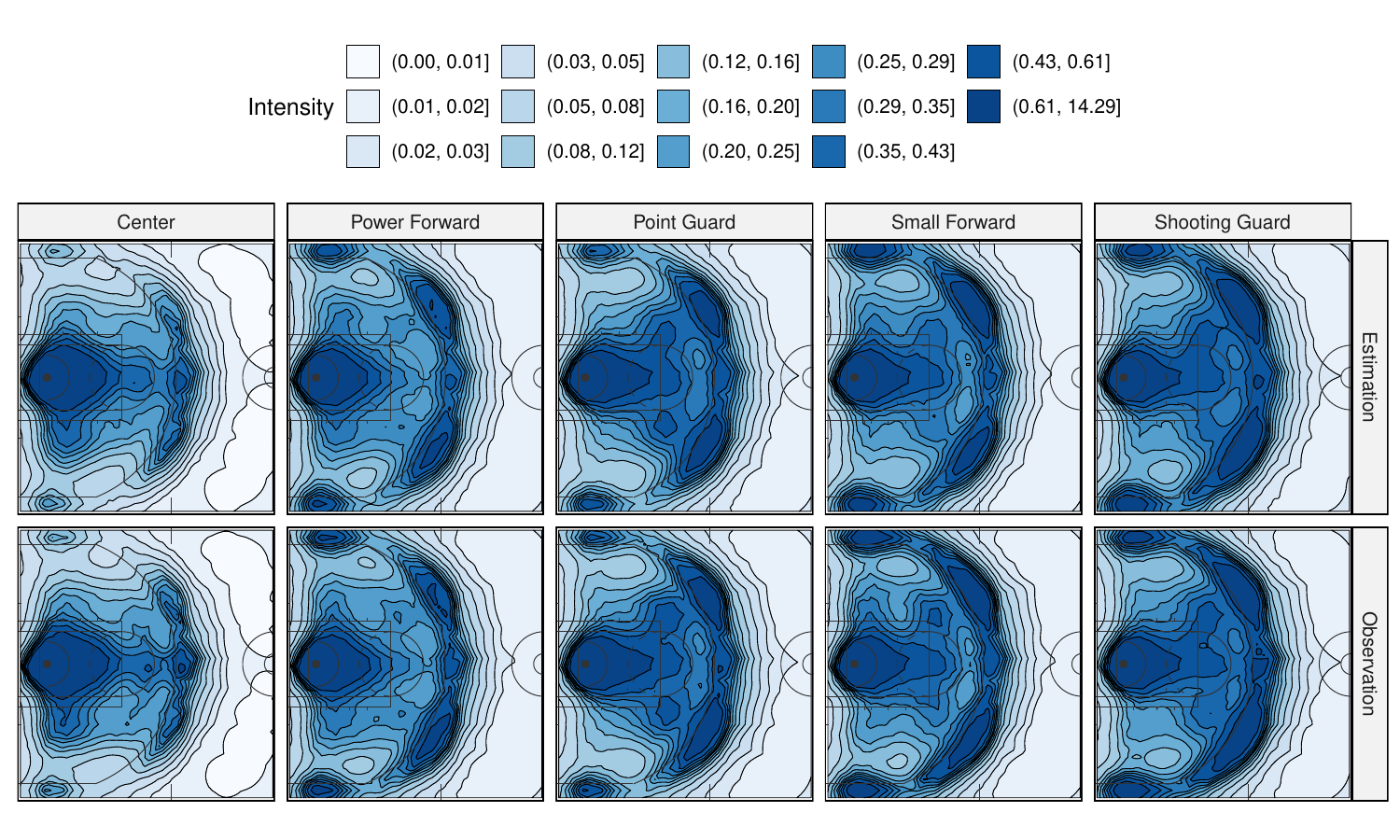}
    \caption{Mean intensity surfaces averaged on players in different positions, and their respective AFBART posterior mean estimates.}
    \label{fig:intensity_surface:avgsurface}
\end{figure}

In addition, we provide one synthetic player with average features for each position, serving as a representative benchmark to analyze and compare performance metrics across different roles on the field. The posterior mean intensity-surface estimates for synthetic players are shown in Figure~\ref{fig:intensity_surface:positionavg}. This approach allows us to evaluate how individual players deviate from the norm, whether overperforming or underperforming relative to their positional peers. Also, these synthetic players can be used to test tactical scenarios, assess the impact of system changes, or even predict how a team might perform if certain positions were filled by players with average capabilities. The posterior mean intensity surfaces averaged on players in different positions are shown in Figure~\ref{fig:intensity_surface:avgsurface}. This offers a comprehensive visualization of their shooting patterns on the field. By analyzing these intensity surfaces, we can identify key areas where players in specific positions tend to shot. Additionally, these estimates allow us to compare the behavior of average players' to elite players, highlighting differences in positioning. This analysis not only enhances our understanding of positional roles but also provides valuable insights for coaching strategies, player development, and opponent scouting.

\begin{figure}[H]
    \centering
    \includegraphics[width=0.8\linewidth]{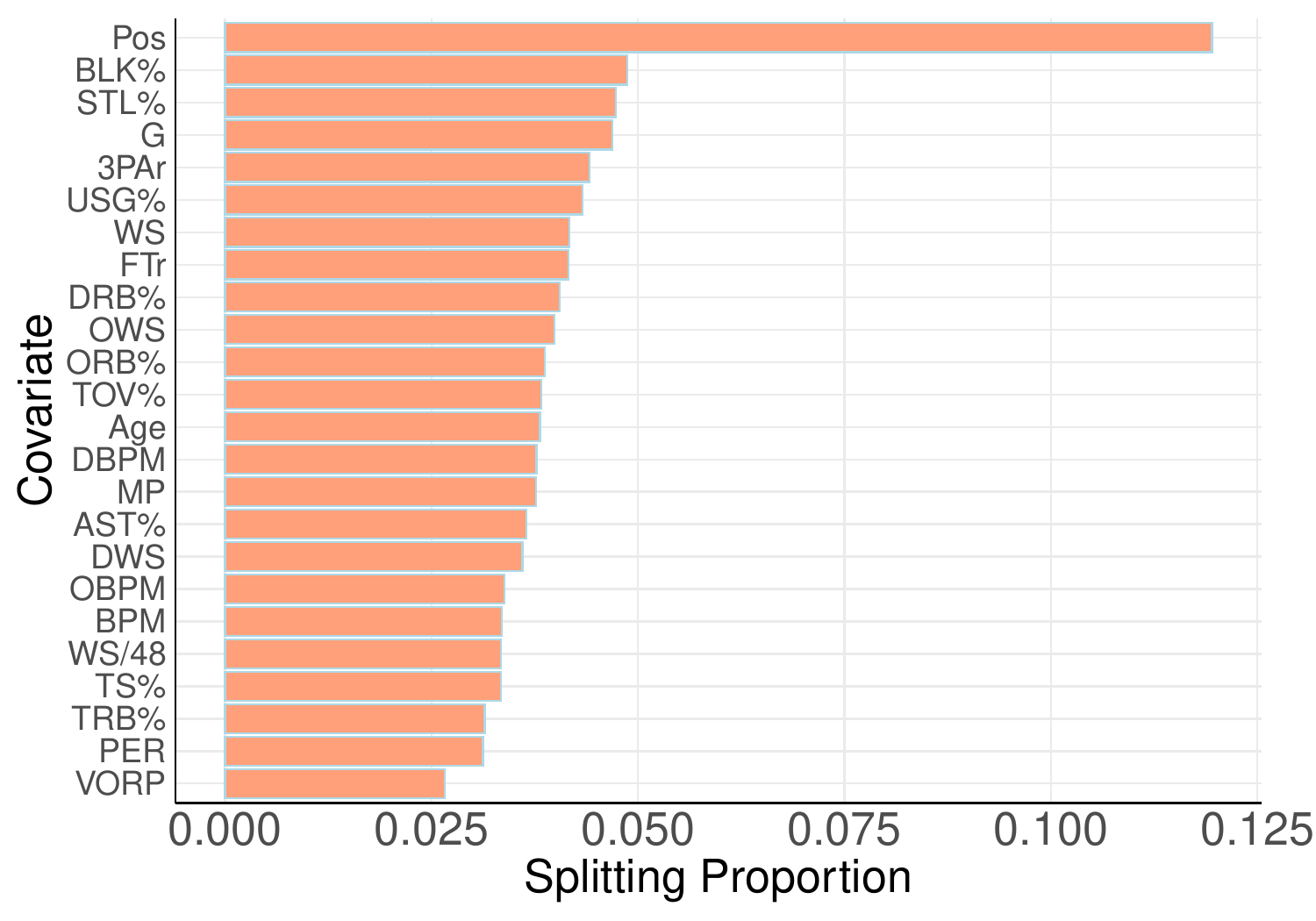}
    \caption{Variable Importance based on splitting proportion.}
    \label{fig:importance}
\end{figure}

AFBART provides measures of variable importance in a fully Bayesian manner. This approach evaluates the contribution of each variable to the model by analyzing how frequently it is used in splitting rules across the ensemble of regression trees. Specifically, for each variable $j \in [p]$ and each tree $t \in [T]$, we define $s_{jt}$ as the proportion of all splitting rules in the $t$-th tree that use the $j$-th component of $\vx$. The splitting proportion $s_j$ of a variable $j$ is defined as the average of $\{s_{jt}\}_{t=1}^T$ across all trees. Intuitively, variables that have a strong influence on the response are more likely to be used in splitting rules, as they help partition the data in a way that improves model fit. Consequently, the splitting proportion $s_j$ serves as a natural and interpretable measure of variable importance. This metric captures the relative contribution of each variable to the model's predictive performance, making it a valuable tool for understanding the underlying structure of the data. Figure~\ref{fig:importance} illustrates the posterior mean splitting proportions for the 24 variables included in the analysis. The results reveal that the top five variables influencing a player's shooting pattern are their position, block percentage, steal percentage, the number of games played, and their 3-point attempt rate. These findings align with basketball intuition, as a player's position dictates their role and typical shot locations, the number of games played reflects their experience, consistency and importance, and the 3-point attempt rate captures their tendency to shoot from beyond the arc. The NBA is currently experiencing a major shift in how the game is played, driven by changes in strategy, player development, and analytics. This transformation named ``three-point revolution" which is particularly evident in the increasing reliance on three-point shooting. Teams and players are prioritizing shots from beyond the arc, as analytics have shown that three-pointers are among the most efficient scoring opportunities in basketball. For example, Joel Embiid is doing things big men never did previously. He had some shot position around the three point arc. This trend has led to a significant evolution in playing styles, with players expanding their shooting range and coaches designing offensive schemes to create more open three-point attempts. As a result, the league has seen a dramatic rise in three-point attempts per game, fundamentally altering the dynamics of the sport. Together, these variables provide key insights into the factors that shape shooting behavior in the dataset.

Finally, we compare the proposed AFBART with two FBART models with fixed basis specifications using a four-fold cross-validation. Specifically, the fixed basis functions of FBART are chosen either as $J=20$ thin plate spline basis functions (denoted as ``FBART-TPS") or as the first $J=20$ functional principal components derived from each training dataset (denoted as ``FBART-FPC"). The performance of these three approaches is evaluated using the average RMSPE and average MCRPS on the respective testing set over 4 cross-validation prediction results. Average RMSPE provides a measure of out-of-sample estimation accuracy, and average MCRPS assesses the quality of out-of-sample uncertainty quantification. The results reveal that AFBART simultaneously achieves the lowest average RMSPE ($0.64$) and the lowest average MCRPS ($0.34$) among all three methods, indicating the best performance in both estimation and uncertainty quantification. In comparison, FBART-TPS yields an average RMSPE of $0.67$ and an average MCRPS of $0.36$, while FBART-FPC yields an average RMSPE of $0.68$ and an average MCRPS of $0.36$. These observations highlight the benefits of adopting an adaptive basis representation when modeling noisy real-world datasets.


\section{Discussions}\label{sec:discuss}
In this paper, we introduce the AFBART model, an advanced framework for analyzing NBA players' spatial intensity maps of field goal attempts. AFBART integrates a reduced-rank basis representation for functional responses with a tree-based approach for basis coefficient functions. It utilizes nonparametric models for both components, effectively tackling the Function-on-Scalars Regression (FOSR) problem by adaptively linking players' intensity surfaces to scalar covariates through data-driven basis functions. The model incorporates regularization priors and a sum-of-trees approach to control complexity and prevent overfitting, enabling the analysis of complex, nonlinear relationships in the data. For efficient computation and inference, we employ a Markov Chain Monte Carlo (MCMC) algorithm, which iteratively samples from the posterior distribution of model parameters to ensure convergence and robust statistical analysis within this hierarchical Bayesian framework.

We applied the AFBART model to shot data from the 2017-2018 NBA regular season. The comparison of posterior mean intensity surfaces with observed shot patterns shows a high degree of similarity, confirming that AFBART accurately captures the spatial distribution of player shot selection. This reliability is further evidenced by creating synthetic players with average features for each position, providing benchmarks for comparing individual players and positional norms. The model offers crucial insights into shooting patterns and strategies, reflecting significant transformations in basketball, particularly the increased emphasis on three-point shooting.

Furthermore, a four-fold cross-validation comparing AFBART to traditional FBART models reveals superior performance in estimation accuracy and uncertainty quantification, highlighting its adaptability to complex, noisy data scenarios typical in sports analytics. This comprehensive analysis not only deepens our understanding of shooting dynamics but also informs coaching strategies and player development.

Several topics beyond the scope of this paper merit further investigation to enhance shooting pattern analysis. First, incorporating covariate measurement error into AFBART could address potential inaccuracies in player characteristics derived from historical records, like three-point attempt rates. Enhancing the basis representation of shot intensity surfaces to include the underlying geometry of the half-court could improve basis efficiency and model interpretability. Additionally, extending AFBART to time series of functional data would allow for analyzing temporal dynamics in shooting patterns, enabling trend identification and change-point detection. Finally, exploring other BART variants within the AFBART framework could expand its capabilities, allowing for more nuanced variable selection and targeted smoothing.

\begin{funding}
This project is partially funded by the Cancer Prevention and Research Institute of Texas (CPRIT) RP230036 and the National Science Foundation (NSF) DMS-2412923/SES-2412922 (Cao and Hu).

\end{funding}

\begin{supplement}
\stitle{Supplement A: Supplementary Material of ``How do the professional players select the shot locations? An analysis of Field Goal Attempts via Bayesian Additive Regression Tree"}
\sdescription{The PDF file contains additional data descriptions, implementation details, and numerical results.}
\end{supplement}


\bibliographystyle{imsart-nameyear} 
\bibliography{ref}       

\begin{thebibliography}{46}

\bibitem[\protect\citeauthoryear{Albert et~al.}{2017}]{albert2017handbook}
\begin{bbook}[author]
\bauthor{\bsnm{Albert},~\bfnm{Jim}\binits{J.}}, \bauthor{\bsnm{Glickman},~\bfnm{Mark~E}\binits{M.~E.}}, \bauthor{\bsnm{Swartz},~\bfnm{Tim~B}\binits{T.~B.}} \AND \bauthor{\bsnm{Koning},~\bfnm{Ruud~H}\binits{R.~H.}}
(\byear{2017}).
\btitle{Handbook of statistical methods and analyses in sports}.
\bpublisher{CRC Press}.
\end{bbook}
\endbibitem

\bibitem[\protect\citeauthoryear{Cao, He and Zhang}{2025}]{cao2025functional}
\begin{barticle}[author]
\bauthor{\bsnm{Cao},~\bfnm{Jiahao}\binits{J.}}, \bauthor{\bsnm{He},~\bfnm{Shiyuan}\binits{S.}} \AND \bauthor{\bsnm{Zhang},~\bfnm{Bohai}\binits{B.}}
(\byear{2025}).
\btitle{Functional Bayesian Additive Regression Trees with Shape Constraints}.
\bjournal{arXiv preprint arXiv:2502.16888}.
\end{barticle}
\endbibitem

\bibitem[\protect\citeauthoryear{Cervone et~al.}{2014}]{cervone2014pointwise}
\begin{binproceedings}[author]
\bauthor{\bsnm{Cervone},~\bfnm{Dan}\binits{D.}}, \bauthor{\bsnm{D’Amour},~\bfnm{Alexander}\binits{A.}}, \bauthor{\bsnm{Bornn},~\bfnm{Luke}\binits{L.}} \AND \bauthor{\bsnm{Goldsberry},~\bfnm{Kirk}\binits{K.}}
(\byear{2014}).
\btitle{POINTWISE: Predicting points and valuing decisions in real time with NBA optical tracking data}.
In \bbooktitle{Proceedings of the 8th MIT Sloan Sports Analytics Conference, Boston, MA, USA}
\bvolume{28}
\bpages{3}.
\end{binproceedings}
\endbibitem

\bibitem[\protect\citeauthoryear{Cervone et~al.}{2016}]{cervone2016multiresolution}
\begin{barticle}[author]
\bauthor{\bsnm{Cervone},~\bfnm{Daniel}\binits{D.}}, \bauthor{\bsnm{D'Amour},~\bfnm{Alex}\binits{A.}}, \bauthor{\bsnm{Bornn},~\bfnm{Luke}\binits{L.}} \AND \bauthor{\bsnm{Goldsberry},~\bfnm{Kirk}\binits{K.}}
(\byear{2016}).
\btitle{A multiresolution stochastic process model for predicting basketball possession outcomes}.
\bjournal{Journal of the American Statistical Association}
\bvolume{111}
\bpages{585--599}.
\end{barticle}
\endbibitem

\bibitem[\protect\citeauthoryear{Chen, Goldsmith and Ogden}{2016}]{chen2016variable}
\begin{barticle}[author]
\bauthor{\bsnm{Chen},~\bfnm{Yakuan}\binits{Y.}}, \bauthor{\bsnm{Goldsmith},~\bfnm{Jeff}\binits{J.}} \AND \bauthor{\bsnm{Ogden},~\bfnm{R~Todd}\binits{R.~T.}}
(\byear{2016}).
\btitle{Variable selection in function-on-scalar regression}.
\bjournal{Stat}
\bvolume{5}
\bpages{88--101}.
\end{barticle}
\endbibitem

\bibitem[\protect\citeauthoryear{Chipman, George and McCulloch}{2010}]{chipman2010bart}
\begin{barticle}[author]
\bauthor{\bsnm{Chipman},~\bfnm{Hugh~A}\binits{H.~A.}}, \bauthor{\bsnm{George},~\bfnm{Edward~I}\binits{E.~I.}} \AND \bauthor{\bsnm{McCulloch},~\bfnm{Robert~E}\binits{R.~E.}}
(\byear{2010}).
\btitle{BART: Bayesian additive regression trees}.
\end{barticle}
\endbibitem

\bibitem[\protect\citeauthoryear{Fan and M{\"u}ller}{2022}]{fan2022conditional}
\begin{barticle}[author]
\bauthor{\bsnm{Fan},~\bfnm{Jianing}\binits{J.}} \AND \bauthor{\bsnm{M{\"u}ller},~\bfnm{Hans-Georg}\binits{H.-G.}}
(\byear{2022}).
\btitle{Conditional distribution regression for functional responses}.
\bjournal{Scandinavian Journal of Statistics}
\bvolume{49}
\bpages{502--524}.
\end{barticle}
\endbibitem

\bibitem[\protect\citeauthoryear{Fernandez and Bornn}{2018}]{fernandez2018wide}
\begin{binproceedings}[author]
\bauthor{\bsnm{Fernandez},~\bfnm{Javier}\binits{J.}} \AND \bauthor{\bsnm{Bornn},~\bfnm{Luke}\binits{L.}}
(\byear{2018}).
\btitle{Wide open spaces: A statistical technique for measuring space creation in professional soccer}.
In \bbooktitle{Sloan Sports Analytics Conference}
\bvolume{2018}.
\end{binproceedings}
\endbibitem

\bibitem[\protect\citeauthoryear{Franks et~al.}{2015}]{franks2015characterizing}
\begin{barticle}[author]
\bauthor{\bsnm{Franks},~\bfnm{Alexander}\binits{A.}}, \bauthor{\bsnm{Miller},~\bfnm{Andrew}\binits{A.}}, \bauthor{\bsnm{Bornn},~\bfnm{Luke}\binits{L.}}, \bauthor{\bsnm{Goldsberry},~\bfnm{Kirk}\binits{K.}} \betal{et~al.}
(\byear{2015}).
\btitle{Characterizing the spatial structure of defensive skill in professional basketball}.
\bjournal{The Annals of Applied Statistics}
\bvolume{9}
\bpages{94--121}.
\end{barticle}
\endbibitem

\bibitem[\protect\citeauthoryear{Ghosal et~al.}{2023}]{ghosal2023shape}
\begin{barticle}[author]
\bauthor{\bsnm{Ghosal},~\bfnm{Rahul}\binits{R.}}, \bauthor{\bsnm{Ghosh},~\bfnm{Sujit}\binits{S.}}, \bauthor{\bsnm{Urbanek},~\bfnm{Jacek}\binits{J.}}, \bauthor{\bsnm{Schrack},~\bfnm{Jennifer~A}\binits{J.~A.}} \AND \bauthor{\bsnm{Zipunnikov},~\bfnm{Vadim}\binits{V.}}
(\byear{2023}).
\btitle{Shape-constrained estimation in functional regression with Bernstein polynomials}.
\bjournal{Computational Statistics \& Data Analysis}
\bvolume{178}
\bpages{107614}.
\end{barticle}
\endbibitem

\bibitem[\protect\citeauthoryear{Gneiting and Raftery}{2007}]{gneiting2007strictly}
\begin{barticle}[author]
\bauthor{\bsnm{Gneiting},~\bfnm{Tilmann}\binits{T.}} \AND \bauthor{\bsnm{Raftery},~\bfnm{Adrian~E}\binits{A.~E.}}
(\byear{2007}).
\btitle{Strictly proper scoring rules, prediction, and estimation}.
\bjournal{Journal of the American Statistical Association}
\bvolume{102}
\bpages{359--378}.
\end{barticle}
\endbibitem

\bibitem[\protect\citeauthoryear{Goldsmith, Greven and Crainiceanu}{2013}]{goldsmith2013corrected}
\begin{barticle}[author]
\bauthor{\bsnm{Goldsmith},~\bfnm{Jeff}\binits{J.}}, \bauthor{\bsnm{Greven},~\bfnm{Sonja}\binits{S.}} \AND \bauthor{\bsnm{Crainiceanu},~\bfnm{CIPRIAN}\binits{C.}}
(\byear{2013}).
\btitle{Corrected confidence bands for functional data using principal components}.
\bjournal{Biometrics}
\bvolume{69}
\bpages{41--51}.
\end{barticle}
\endbibitem

\bibitem[\protect\citeauthoryear{Greven et~al.}{2011}]{greven2011longitudinal}
\begin{binproceedings}[author]
\bauthor{\bsnm{Greven},~\bfnm{Sonja}\binits{S.}}, \bauthor{\bsnm{Crainiceanu},~\bfnm{Ciprian}\binits{C.}}, \bauthor{\bsnm{Caffo},~\bfnm{Brian}\binits{B.}} \AND \bauthor{\bsnm{Reich},~\bfnm{Daniel}\binits{D.}}
(\byear{2011}).
\btitle{Longitudinal functional principal component analysis}.
In \bbooktitle{Recent Advances in Functional Data Analysis and Related Topics}
\bpages{149--154}.
\bpublisher{Springer}.
\end{binproceedings}
\endbibitem

\bibitem[\protect\citeauthoryear{Hahn, Murray and Carvalho}{2020}]{hahn2020bayesian}
\begin{barticle}[author]
\bauthor{\bsnm{Hahn},~\bfnm{P~Richard}\binits{P.~R.}}, \bauthor{\bsnm{Murray},~\bfnm{Jared~S}\binits{J.~S.}} \AND \bauthor{\bsnm{Carvalho},~\bfnm{Carlos~M}\binits{C.~M.}}
(\byear{2020}).
\btitle{Bayesian regression tree models for causal inference: Regularization, confounding, and heterogeneous effects (with discussion)}.
\bjournal{Bayesian Analysis}
\bvolume{15}
\bpages{965--1056}.
\end{barticle}
\endbibitem

\bibitem[\protect\citeauthoryear{Hill, Linero and Murray}{2020}]{hill2020bayesian}
\begin{barticle}[author]
\bauthor{\bsnm{Hill},~\bfnm{Jennifer}\binits{J.}}, \bauthor{\bsnm{Linero},~\bfnm{Antonio}\binits{A.}} \AND \bauthor{\bsnm{Murray},~\bfnm{Jared}\binits{J.}}
(\byear{2020}).
\btitle{Bayesian additive regression trees: {A} review and look forward}.
\bjournal{Annual Review of Statistics and Its Application}
\bvolume{7}
\bpages{251--278}.
\end{barticle}
\endbibitem

\bibitem[\protect\citeauthoryear{Hu, Yang and Xue}{2021}]{hu2021bayesian}
\begin{barticle}[author]
\bauthor{\bsnm{Hu},~\bfnm{Guanyu}\binits{G.}}, \bauthor{\bsnm{Yang},~\bfnm{Hou-Cheng}\binits{H.-C.}} \AND \bauthor{\bsnm{Xue},~\bfnm{Yishu}\binits{Y.}}
(\byear{2021}).
\btitle{Bayesian group learning for shot selection of professional basketball players}.
\bjournal{Stat}
\bvolume{10}
\bpages{e324}.
\end{barticle}
\endbibitem

\bibitem[\protect\citeauthoryear{Hu et~al.}{2022}]{hu2021cjs}
\begin{barticle}[author]
\bauthor{\bsnm{Hu},~\bfnm{Guanyu}\binits{G.}}, \bauthor{\bsnm{Yang},~\bfnm{Hou-Cheng}\binits{H.-C.}}, \bauthor{\bsnm{Xue},~\bfnm{Yishu}\binits{Y.}} \AND \bauthor{\bsnm{Dey},~\bfnm{Dipak~K.}\binits{D.~K.}}
(\byear{2022}).
\btitle{Zero-inflated Poisson model with clustered regression coefficients: Application to heterogeneity learning of field goal attempts of professional basketball players}.
\bjournal{Canadian Journal of Statistics}.
\end{barticle}
\endbibitem

\bibitem[\protect\citeauthoryear{Huang and Lin}{2020}]{huang2020regression}
\begin{barticle}[author]
\bauthor{\bsnm{Huang},~\bfnm{Mei-Ling}\binits{M.-L.}} \AND \bauthor{\bsnm{Lin},~\bfnm{Yi-Jung}\binits{Y.-J.}}
(\byear{2020}).
\btitle{Regression tree model for predicting game scores for the golden state warriors in the national basketball association}.
\bjournal{Symmetry}
\bvolume{12}
\bpages{835}.
\end{barticle}
\endbibitem

\bibitem[\protect\citeauthoryear{Jiao, Hu and Yan}{2021}]{jiao2019bayesian}
\begin{barticle}[author]
\bauthor{\bsnm{Jiao},~\bfnm{Jieying}\binits{J.}}, \bauthor{\bsnm{Hu},~\bfnm{Guanyu}\binits{G.}} \AND \bauthor{\bsnm{Yan},~\bfnm{Jun}\binits{J.}}
(\byear{2021}).
\btitle{A {B}ayesian marked spatial point processes model for basketball shot chart}.
\bjournal{Journal of Quantitative Analysis in Sports}
\bvolume{17}
\bpages{77--90}.
\end{barticle}
\endbibitem

\bibitem[\protect\citeauthoryear{Kowal}{2018}]{kowal2018dynamic}
\begin{barticle}[author]
\bauthor{\bsnm{Kowal},~\bfnm{Daniel~R}\binits{D.~R.}}
(\byear{2018}).
\btitle{Dynamic function-on-scalars regression}.
\bjournal{arXiv preprint arXiv:1806.01460}.
\end{barticle}
\endbibitem

\bibitem[\protect\citeauthoryear{Kowal and Bourgeois}{2020}]{kowal2020bayesian}
\begin{barticle}[author]
\bauthor{\bsnm{Kowal},~\bfnm{Daniel~R}\binits{D.~R.}} \AND \bauthor{\bsnm{Bourgeois},~\bfnm{Daniel~C}\binits{D.~C.}}
(\byear{2020}).
\btitle{Bayesian function-on-scalars regression for high-dimensional data}.
\bjournal{Journal of Computational and Graphical Statistics}
\bvolume{29}
\bpages{629--638}.
\end{barticle}
\endbibitem

\bibitem[\protect\citeauthoryear{Li, Linero and Murray}{2023}]{li2023adaptive}
\begin{barticle}[author]
\bauthor{\bsnm{Li},~\bfnm{Yinpu}\binits{Y.}}, \bauthor{\bsnm{Linero},~\bfnm{Antonio~R}\binits{A.~R.}} \AND \bauthor{\bsnm{Murray},~\bfnm{Jared}\binits{J.}}
(\byear{2023}).
\btitle{Adaptive conditional distribution estimation with Bayesian decision tree ensembles}.
\bjournal{Journal of the American Statistical Association}
\bvolume{118}
\bpages{2129--2142}.
\end{barticle}
\endbibitem

\bibitem[\protect\citeauthoryear{Linero and Yang}{2018}]{linero2018bayesian}
\begin{barticle}[author]
\bauthor{\bsnm{Linero},~\bfnm{Antonio~R}\binits{A.~R.}} \AND \bauthor{\bsnm{Yang},~\bfnm{Yun}\binits{Y.}}
(\byear{2018}).
\btitle{Bayesian regression tree ensembles that adapt to smoothness and sparsity}.
\bjournal{Journal of the Royal Statistical Society Series B: Statistical Methodology}
\bvolume{80}
\bpages{1087--1110}.
\end{barticle}
\endbibitem

\bibitem[\protect\citeauthoryear{Luo, Sang and Mallick}{2021}]{luo2021bast}
\begin{barticle}[author]
\bauthor{\bsnm{Luo},~\bfnm{Zhao~Tang}\binits{Z.~T.}}, \bauthor{\bsnm{Sang},~\bfnm{Huiyan}\binits{H.}} \AND \bauthor{\bsnm{Mallick},~\bfnm{Bani}\binits{B.}}
(\byear{2021}).
\btitle{BAST: Bayesian additive regression spanning trees for complex constrained domain}.
\bjournal{Advances in Neural Information Processing Systems}
\bvolume{34}
\bpages{90--102}.
\end{barticle}
\endbibitem

\bibitem[\protect\citeauthoryear{Miller et~al.}{2014}]{miller2014factorized}
\begin{binproceedings}[author]
\bauthor{\bsnm{Miller},~\bfnm{Andrew}\binits{A.}}, \bauthor{\bsnm{Bornn},~\bfnm{Luke}\binits{L.}}, \bauthor{\bsnm{Adams},~\bfnm{Ryan}\binits{R.}} \AND \bauthor{\bsnm{Goldsberry},~\bfnm{Kirk}\binits{K.}}
(\byear{2014}).
\btitle{Factorized point process intensities: A spatial analysis of professional basketball}.
In \bbooktitle{International conference on machine learning}
\bpages{235--243}.
\end{binproceedings}
\endbibitem

\bibitem[\protect\citeauthoryear{M{\o}ller, Syversveen and Waagepetersen}{1998}]{moller1998log}
\begin{barticle}[author]
\bauthor{\bsnm{M{\o}ller},~\bfnm{Jesper}\binits{J.}}, \bauthor{\bsnm{Syversveen},~\bfnm{Anne~Randi}\binits{A.~R.}} \AND \bauthor{\bsnm{Waagepetersen},~\bfnm{Rasmus~Plenge}\binits{R.~P.}}
(\byear{1998}).
\btitle{Log gaussian cox processes}.
\bjournal{Scandinavian journal of statistics}
\bvolume{25}
\bpages{451--482}.
\end{barticle}
\endbibitem

\bibitem[\protect\citeauthoryear{Morris}{2015}]{morris2015functional}
\begin{barticle}[author]
\bauthor{\bsnm{Morris},~\bfnm{Jeffrey~S}\binits{J.~S.}}
(\byear{2015}).
\btitle{Functional regression}.
\bjournal{Annual Review of Statistics and Its Application}
\bvolume{2}
\bpages{321--359}.
\end{barticle}
\endbibitem

\bibitem[\protect\citeauthoryear{Morris and Carroll}{2006}]{morris2006wavelet}
\begin{barticle}[author]
\bauthor{\bsnm{Morris},~\bfnm{Jeffrey~S}\binits{J.~S.}} \AND \bauthor{\bsnm{Carroll},~\bfnm{Raymond~J}\binits{R.~J.}}
(\byear{2006}).
\btitle{Wavelet-based functional mixed models}.
\bjournal{Journal of the Royal Statistical Society Series B: Statistical Methodology}
\bvolume{68}
\bpages{179--199}.
\end{barticle}
\endbibitem

\bibitem[\protect\citeauthoryear{Petersen and M{\"u}ller}{2019}]{petersen2019frechet}
\begin{barticle}[author]
\bauthor{\bsnm{Petersen},~\bfnm{Alexander}\binits{A.}} \AND \bauthor{\bsnm{M{\"u}ller},~\bfnm{Hans-Georg}\binits{H.-G.}}
(\byear{2019}).
\btitle{Fr{\'e}chet regression for random objects with Euclidean predictors}.
\bjournal{The Annals of Statistics}
\bvolume{47}
\bpages{691--719}.
\end{barticle}
\endbibitem

\bibitem[\protect\citeauthoryear{Qi, Hu and Wu}{2024}]{qi2025aoas}
\begin{barticle}[author]
\bauthor{\bsnm{Qi},~\bfnm{Kai}\binits{K.}}, \bauthor{\bsnm{Hu},~\bfnm{Guanyu}\binits{G.}} \AND \bauthor{\bsnm{Wu},~\bfnm{Wei}\binits{W.}}
(\byear{2024}).
\btitle{{Are made and missed different? An analysis of field goal attempts of professional basketball players via depth based testing procedure}}.
\bjournal{The Annals of Applied Statistics}
\bvolume{18}
\bpages{2615 -- 2634}.
\bdoi{10.1214/24-AOAS1899}
\end{barticle}
\endbibitem

\bibitem[\protect\citeauthoryear{Ramsay and Dalzell}{1991}]{ramsay1991some}
\begin{barticle}[author]
\bauthor{\bsnm{Ramsay},~\bfnm{J.~O.}\binits{J.~O.}} \AND \bauthor{\bsnm{Dalzell},~\bfnm{C.~J.}\binits{C.~J.}}
(\byear{1991}).
\btitle{Some Tools for Functional Data Analysis}.
\bjournal{Journal of the Royal Statistical Society: Series B (Methodological)}
\bvolume{53}
\bpages{539--561}.
\end{barticle}
\endbibitem

\bibitem[\protect\citeauthoryear{Ramsay and Silverman}{2005}]{ramsay2005functional}
\begin{bbook}[author]
\bauthor{\bsnm{Ramsay},~\bfnm{J.~O.}\binits{J.~O.}} \AND \bauthor{\bsnm{Silverman},~\bfnm{B.~W.}\binits{B.~W.}}
(\byear{2005}).
\btitle{Functional Data Analysis},
\bedition{2} ed.
\bseries{Springer Series in Statistics}.
\bpublisher{Springer New York, NY}
\bnote{Published: 08 June 2005, Softcover Published: 10 November 2010, eBook Published: 28 June 2006}.
\end{bbook}
\endbibitem

\bibitem[\protect\citeauthoryear{Reich et~al.}{2006}]{reich2006spatial}
\begin{barticle}[author]
\bauthor{\bsnm{Reich},~\bfnm{Brian~J}\binits{B.~J.}}, \bauthor{\bsnm{Hodges},~\bfnm{James~S}\binits{J.~S.}}, \bauthor{\bsnm{Carlin},~\bfnm{Bradley~P}\binits{B.~P.}} \AND \bauthor{\bsnm{Reich},~\bfnm{Adam~M}\binits{A.~M.}}
(\byear{2006}).
\btitle{A spatial analysis of basketball shot chart data}.
\bjournal{The American Statistician}
\bvolume{60}
\bpages{3--12}.
\end{barticle}
\endbibitem

\bibitem[\protect\citeauthoryear{Rosen and Thompson}{2009}]{rosen2009bayesian}
\begin{barticle}[author]
\bauthor{\bsnm{Rosen},~\bfnm{Ori}\binits{O.}} \AND \bauthor{\bsnm{Thompson},~\bfnm{Wesley~K}\binits{W.~K.}}
(\byear{2009}).
\btitle{A Bayesian regression model for multivariate functional data}.
\bjournal{Computational statistics \& data analysis}
\bvolume{53}
\bpages{3773--3786}.
\end{barticle}
\endbibitem

\bibitem[\protect\citeauthoryear{Ro{\v{c}}kov{\'a} and Saha}{2019}]{rovckova2019theory}
\begin{binproceedings}[author]
\bauthor{\bsnm{Ro{\v{c}}kov{\'a}},~\bfnm{Veronika}\binits{V.}} \AND \bauthor{\bsnm{Saha},~\bfnm{Enakshi}\binits{E.}}
(\byear{2019}).
\btitle{On theory for BART}.
In \bbooktitle{The 22nd international conference on artificial intelligence and statistics}
\bpages{2839--2848}.
\bpublisher{PMLR}.
\end{binproceedings}
\endbibitem

\bibitem[\protect\citeauthoryear{Ro{\v{c}}kov{\'a} and Van~der Pas}{2020}]{rovckova2020posterior}
\begin{barticle}[author]
\bauthor{\bsnm{Ro{\v{c}}kov{\'a}},~\bfnm{Veronika}\binits{V.}} \AND \bauthor{\bparticle{Van~der} \bsnm{Pas},~\bfnm{Stephanie}\binits{S.}}
(\byear{2020}).
\btitle{Posterior concentration for Bayesian regression trees and forests}.
\bjournal{The Annals of Statistics}
\bvolume{48}
\bpages{2108--2131}.
\end{barticle}
\endbibitem

\bibitem[\protect\citeauthoryear{Sandholtz, Mortensen and Bornn}{2020}]{sandholtz2019measuring}
\begin{barticle}[author]
\bauthor{\bsnm{Sandholtz},~\bfnm{Nathan}\binits{N.}}, \bauthor{\bsnm{Mortensen},~\bfnm{Jacob}\binits{J.}} \AND \bauthor{\bsnm{Bornn},~\bfnm{Luke}\binits{L.}}
(\byear{2020}).
\btitle{Measuring spatial allocative efficiency in basketball}.
\bjournal{Journal of Quantitative Analysis in Sports}
\bvolume{16}
\bpages{271--289}.
\end{barticle}
\endbibitem

\bibitem[\protect\citeauthoryear{Starling et~al.}{2020}]{starling2020bart}
\begin{barticle}[author]
\bauthor{\bsnm{Starling},~\bfnm{Jennifer~E}\binits{J.~E.}}, \bauthor{\bsnm{Murray},~\bfnm{Jared~S}\binits{J.~S.}}, \bauthor{\bsnm{Carvalho},~\bfnm{Carlos~M}\binits{C.~M.}}, \bauthor{\bsnm{Bukowski},~\bfnm{Radek~K}\binits{R.~K.}} \AND \bauthor{\bsnm{Scott},~\bfnm{James~G}\binits{J.~G.}}
(\byear{2020}).
\btitle{BART WITH TARGETED SMOOTHING}.
\bjournal{The Annals of Applied Statistics}
\bvolume{14}
\bpages{28--50}.
\end{barticle}
\endbibitem

\bibitem[\protect\citeauthoryear{Wang, Chiou and M{\"u}ller}{2016}]{wang2016functional}
\begin{barticle}[author]
\bauthor{\bsnm{Wang},~\bfnm{Jane-Ling}\binits{J.-L.}}, \bauthor{\bsnm{Chiou},~\bfnm{Jeng-Min}\binits{J.-M.}} \AND \bauthor{\bsnm{M{\"u}ller},~\bfnm{Hans-Georg}\binits{H.-G.}}
(\byear{2016}).
\btitle{Functional data analysis}.
\bjournal{Annual Review of Statistics and its application}
\bvolume{3}
\bpages{257--295}.
\end{barticle}
\endbibitem

\bibitem[\protect\citeauthoryear{Wong-Toi et~al.}{2023}]{wong2023joint}
\begin{barticle}[author]
\bauthor{\bsnm{Wong-Toi},~\bfnm{Eliot}\binits{E.}}, \bauthor{\bsnm{Yang},~\bfnm{Hou-Cheng}\binits{H.-C.}}, \bauthor{\bsnm{Shen},~\bfnm{Weining}\binits{W.}} \AND \bauthor{\bsnm{Hu},~\bfnm{Guanyu}\binits{G.}}
(\byear{2023}).
\btitle{A Joint Analysis for Field Goal Attempts and Percentages of Professional Basketball Players: Bayesian Nonparametric Resource.}
\bjournal{Journal of Data Science}
\bvolume{21}.
\end{barticle}
\endbibitem

\bibitem[\protect\citeauthoryear{Wood}{2017}]{wood2017generalized}
\begin{bbook}[author]
\bauthor{\bsnm{Wood},~\bfnm{Simon~N}\binits{S.~N.}}
(\byear{2017}).
\btitle{Generalized additive models: an introduction with R}.
\bpublisher{chapman and hall/CRC}.
\end{bbook}
\endbibitem

\bibitem[\protect\citeauthoryear{Wu and Bornn}{2018}]{wu2018modeling}
\begin{barticle}[author]
\bauthor{\bsnm{Wu},~\bfnm{Steven}\binits{S.}} \AND \bauthor{\bsnm{Bornn},~\bfnm{Luke}\binits{L.}}
(\byear{2018}).
\btitle{Modeling offensive player movement in professional basketball}.
\bjournal{The American Statistician}
\bvolume{72}
\bpages{72--79}.
\end{barticle}
\endbibitem

\bibitem[\protect\citeauthoryear{Yee, Ghosh and Deshpande}{2024}]{yee2024scalable}
\begin{barticle}[author]
\bauthor{\bsnm{Yee},~\bfnm{Ryan}\binits{R.}}, \bauthor{\bsnm{Ghosh},~\bfnm{Soham}\binits{S.}} \AND \bauthor{\bsnm{Deshpande},~\bfnm{Sameer~K}\binits{S.~K.}}
(\byear{2024}).
\btitle{Scalable piecewise smoothing with BART}.
\bjournal{arXiv preprint arXiv:2411.07984}.
\end{barticle}
\endbibitem

\bibitem[\protect\citeauthoryear{Yin, Hu and Shen}{2023}]{yin2023analysis}
\begin{barticle}[author]
\bauthor{\bsnm{Yin},~\bfnm{Fan}\binits{F.}}, \bauthor{\bsnm{Hu},~\bfnm{Guanyu}\binits{G.}} \AND \bauthor{\bsnm{Shen},~\bfnm{Weining}\binits{W.}}
(\byear{2023}).
\btitle{Analysis of professional basketball field goal attempts via a Bayesian matrix clustering approach}.
\bjournal{Journal of Computational and Graphical Statistics}
\bvolume{32}
\bpages{49--60}.
\end{barticle}
\endbibitem

\bibitem[\protect\citeauthoryear{Yin et~al.}{2022}]{yin2022bayesian}
\begin{binproceedings}[author]
\bauthor{\bsnm{Yin},~\bfnm{Fan}\binits{F.}}, \bauthor{\bsnm{Jiao},~\bfnm{Jieying}\binits{J.}}, \bauthor{\bsnm{Yan},~\bfnm{Jun}\binits{J.}} \AND \bauthor{\bsnm{Hu},~\bfnm{Guanyu}\binits{G.}}
(\byear{2022}).
\btitle{Bayesian nonparametric learning for point processes with spatial homogeneity: A spatial analysis of NBA shot locations}.
In \bbooktitle{International Conference on Machine Learning}
\bpages{25523--25551}.
\bpublisher{PMLR}.
\end{binproceedings}
\endbibitem

\bibitem[\protect\citeauthoryear{Zuccolotto, Sandri and Manisera}{2023}]{zuccolotto2023spatial}
\begin{barticle}[author]
\bauthor{\bsnm{Zuccolotto},~\bfnm{Paola}\binits{P.}}, \bauthor{\bsnm{Sandri},~\bfnm{Marco}\binits{M.}} \AND \bauthor{\bsnm{Manisera},~\bfnm{Marica}\binits{M.}}
(\byear{2023}).
\btitle{Spatial performance analysis in basketball with CART, random forest and extremely randomized trees}.
\bjournal{Annals of Operations Research}
\bvolume{325}
\bpages{495--519}.
\end{barticle}
\endbibitem

\end{thebibliography}


\end{document}